\documentclass[fleqn,usenatbib]{mnras}

\usepackage{newtxtext,newtxmath}
\usepackage[T1]{fontenc}

\DeclareRobustCommand{\VAN}[3]{#2}
\let\VANthebibliography\thebibliography
\def\thebibliography{\DeclareRobustCommand{\VAN}[3]{##3}\VANthebibliography}

\usepackage{graphicx}	% Including figure files
\usepackage{amsmath}	% Advanced maths commands

\usepackage{bm}	% Allow bold greek letters in math mode (command \bm)

\usepackage[export]{adjustbox}

\newcommand{\noopsort}[1]{}

\newcommand{\teff}{T_{\text{eff}}}

\title[DZ planetesimals I: chondritic compositions]{Planetesimals at DZ stars I: chondritic compositions and a massive accretion event}

\author[A. Swan et al.]
{Andrew Swan$^{1,2}$\thanks{E-mail: \href{mailto:Andrew.Swan@warwick.ac.uk}{Andrew.Swan@warwick.ac.uk}},
Jay Farihi$^{2}$,
Carl Melis$^{3}$,
Patrick Dufour$^{4,5}$,
Steven J. Desch$^{6}$,
Detlev Koester$^{7}$,
\newauthor
and Jincheng Guo$^{2,8}$
\\
$^{1}$Department of Physics, University of Warwick, Coventry~CV4~7AL, UK\\
$^{2}$Department of Physics~\& Astronomy, University College London, Gower Street, London~WC1E~6BT, UK\\
$^{3}$Center for Astrophysics and Space Sciences, University of California, San Diego, CA~92093-0424, USA\\
$^{4}$D\'epartement de Physique, Universit\'e de Montr\'eal, C.P.~6128 Succ. Centre-ville, Montr\'eal, QC~H3C~3J7, Canada\\
$^{5}$Trottier Institute for Research on Exoplanets, Universit\'e de Montr\'eal, C.P.~6128 Succ. Centre-ville, Montr\'eal, QC~H3C~3J7, Canada\\
$^{6}$School of Earth and Space Exploration, Arizona State University, Tempe, AZ 85287-1404, USA\\
$^{7}$Institut f\"ur Theoretische Physik und Astrophysik, University of Kiel, D-24098~Kiel, Germany\\
$^{8}$Department of Scientific Research, Beijing Planetarium, Xizhimenwai Street, Beijing 100044, China\\
}

\date{Accepted XXX. Received YYY; in original form ZZZ}

\pubyear{2023}

\begin{document}
\label{firstpage}
\pagerange{\pageref{firstpage}--\pageref{lastpage}}
\maketitle

%It should be a single paragraph not more than 250 words (200 words for Letters).
\begin{abstract}
There is a wealth of evidence to suggest that planetary systems can survive beyond the main sequence. Most commonly, white dwarfs are found to be accreting material from tidally disrupted asteroids, whose bulk compositions are reflected by the metals polluting the stellar photospheres. While many examples are known, most lack the deep, high-resolution data required to detect multiple elements, and thus characterise the planetesimals that orbit them. Here, spectra of seven DZ~white dwarfs observed with Keck HIRES are analysed, where up to nine metals are measured per star. Their compositions are compared against those of solar system objects, working in a Bayesian framework to infer or marginalise over the accretion history. All of the stars have been accreting primitive material, similar to chondrites, with hints of a Mercury-like composition at one star. The most polluted star is observed several Myr after its last major accretion episode, in which a Moon-sized object met its demise.
\end{abstract}


%%%%%%%%%%%%%%%%%%%%%%%%%%%%%%%%%%%%%%%%%%%%%%%%%%

%%%%%%%%%%%%%%%%% BODY OF PAPER %%%%%%%%%%%%%%%%%%

\section{Introduction}
\label{sectionIntroduction}

Three decades after confirmation of the first exoplanet \citep{Wolszczan1992}, the field has progressed from discovery and cataloguing into an era of characterisation. Instrumental precision has reached the point where detection of Earth-sized worlds is feasible, and the confounding effect of stellar activity is now the main challenge to overcome \citep{Gillon2017,Dumusque2018}. Where a radius and mass are both measured for a planet, and clouds do not obscure the surface, its density can be calculated. From that, planet formation models aim to infer its interior structure and bulk composition. \citep{Unterborn2018}. However, even in the solar system, planetary structures cannot be fully constrained \citep{Shah2022}.

Interior models for terrestrial exoplanets can benefit from knowledge of the host star chemistry, which reflects the primordial composition of the protoplanetary nebula \citep{Dorn2015, Adibekyan2021}. While a wide range of outcomes are possible depending on the formation chemistry \citep{Bond2010}, analysis of solar twins finds little compositional variation between systems \citep{Bedell2018}. Rocky exoplanets may therefore be similar to those in the solar system, motivating searches for an Earth analogue.

Complementary to conventional exoplanet studies are analyses of the planetary systems that survive around white dwarfs. While the surfaces of stellar remnants retain no memory of their main-sequence chemistry, they can give access to the bulk compositions of the objects that formed around them early in their history. This is made possible by the behaviour of metals in white dwarf atmospheres: as the star cools below around 25\,000\,K, radiation pressure is no longer able to support metals against gravity \citep{Chayer1995}. Any metals sink out of the photosphere on timescales in the range $10^{-2}$--$10^7$\,yr, always much shorter than the cooling age of the star \citep{Schatzman1945}. As white dwarf atmospheres are expected to contain only hydrogen or helium, the presence of metal lines in their spectra is a clear signpost of the external delivery of material from their planetary systems. Up to 50~per~cent of systems display such evidence \citep{Koester2014}, corroborated in some cases by optical transits \citep{vanderburg2015,Guidry2021,Farihi2022}, or infrared excesses from circumstellar dust \citep{Zuckerman1987,Jura2007Spitzer11stars,Jura2009silicates,Farihi2016, Wilson2019}.

Over 1000 metal-polluted white dwarfs are known \citep{Coutu2019}, although only a few dozen have been studied in detail. High-resolution spectroscopy on 8--10\,m telescopes, supplemented by ultraviolet observations from space, has resulted in detections of 23 distinct metal species to date, with up to 16 in a single star \citep{Zuckerman2007,Xu2013,Klein2021}. The majority of stars surveyed have abundances consistent with the accretion of rocky, chondritic material \citep{Gansicke2012, Hollands2018analysis, Swan2019Xshooter, Xu2019composition, Trierweiler2023}, but in a few instances volatile-bearing objects have been found \citep{Farihi2013, Raddi2015, Xu2017, Hoskin2020}, including exomoon candidates \citep{Klein2021,Doyle2021} and an ice giant \citep{Gansicke2019}. As well as a spread of compositions, polluted white dwarfs sample a range of ages, with some planet hosts older than the solar system \citep{Hollands2021alkali, Elms2022}.

Such rich datasets have enabled geochemical studies of exoplanets and their building blocks. It has been found that differentiation of planetesimals by radiogenic heating is a ubiquitous process \citep{Jura2013,Bonsor2022,Curry2022}, and inferred oxygen fugacities have been used to show that most rocky objects polluting white dwarfs formed under oxidising conditions similar to those in the early inner solar system \citep{Doyle2019,Doyle2020}. Methods have been developed to determine the internal pressures under which core--mantle differentiation has taken place, and thus the mass of the parent object \citep{Buchan2022}, and to search for evidence of post-nebular volatilisation \citep{Harrison2021postNebula}. Such studies validate the application of familiar geochemistry to other planetary systems, although there are indications that the range of exoplanetary rock compositions exceeds that found on Earth \citep{Putirka2021}.

Among metal-enriched white dwarfs, the DZ~stars tend to dominate the population. Their cool helium atmospheres are sufficiently transparent that abundances as low as $\log{(\text{Ca}/\text{He})}=-10$ can produce strong \ion{Ca}{ii}~H and~K lines. Indeed, such Ca absorption was detected in vMa\,2 using photographic plate technology over a century ago \citep{vanMaanen1917}. Those lines are less prominent in DAZ and DBZ spectra, as seen at hydrogen-rich stars, and in warmer helium atmospheres. In this sense, the DZ~stars may represent the most abundant and readily detectable signposts of exoplanetary systems; in fact, two of the three nearest white dwarfs -- Procyon B, and the prototype vMa\,2 -- are metal-polluted and likely host remnant planetary systems \citep{Farihi2013Hubble}. Hundreds of DZ~stars have been discovered in the Sloan Digital Sky Survey (SDSS; \citealt{York2000SDSS, Eisenstein2006, Dufour2007, Hollands2017abundances}), and have played a pivotal role in the development of the field. Many reside well out of the Galactic plane, far from the dense regions of the interstellar medium that were once thought to be the source of white dwarf pollution, thus decisively implicating an origin from planetary material instead \citep{Zuckerman2010, Farihi2010rockyPlanetesimals}. 

It is important to obtain as many abundances as possible when analysing polluted white dwarfs, as the photospheric metals do not necessarily reflect those in the accreted material. Each element sinks out of the convection zone at a different rate, and this effect must be taken into account \citep{Swan2019Xshooter, Harrison2021bayesian}. For example, excess oxygen beyond that needed to bind to the major rock-forming elements may derive from volatiles such as water ice \citep{Klein2010}. However, an apparent excess may develop even after dry rock has been accreted, as most metals sink away faster than oxygen.

This study presents high-resolution spectra of seven DZ~stars, with at least five metals detected at each target. The accretion process is modelled in a Bayesian framework in order to constrain the effects of differential sinking. Observations and photospheric abundance determination using white dwarf atmosphere models are described in Section~\ref{sectionObservationsAndModelling}. Those data are analysed by comparison to solar system compositions, with methods outlined in Section~\ref{sectionAnalysis} and results presented in Section~\ref{sectionResults}. A discussion follows in Section~\ref{sectionDiscussion}, and conclusions are given in Section~\ref{sectionConclusions}. A full description of the Bayesian model follows in Appendix~\ref{appendixBayesian}.

\section{Observations and stellar modelling}
\label{sectionObservationsAndModelling}

Targets were selected from DZ~stars identified from SDSS, favouring the brightest objects, or those where high $\text{Ca}/\text{He}$ abundances had been measured \citep{Dufour2007}. They were observed with the High Resolution Echelle Spectrometer (HIRES; \citealt{Vogt1994}) on Keck\,I on the nights of 2009~May~23 and~24, using the blue cross-disperser on the first night and the red cross-disperser on the second. Sky conditions were good on May~23, but there were some thin clouds the following night. Spectra covering 3120--5950\,{\AA} and 4720--9000\,{\AA} were obtained, using the 1\farcs15 slit for a resolving power $R\approx37$\,000. Two or three exposures of 900--1200\,s were taken for blue spectra, depending on target brightness, for a total integration time of 1800--3600\,s per star. One or two exposures of 900 or 1800\,s were taken for red spectra, for a total integration time of 900--3600\,s per star. No red data were obtained for 084857.88+002834.9, so the HIRES data are analysed jointly with SDSS spectra for that star.

Two stars (080537.64+383212.4 and 153129.26+424015.7) showed only \ion{Ca}{ii} H and~K lines in their spectra, with $\log{(\text{Ca}/\text{He})}\lesssim-9.5$, and are not analysed further. Seven stars remain in the sample, identified in Table~\ref{tableTargetData}. For brevity, abbreviated SDSS names will be used henceforth.

\begin{table*}
\begin{center}
\caption{Keck/HIRES targets analysed in detail. \textit{Gaia}~DR3 astrometry is used to propagate right ascension and declination to J2000 epoch, and distances are estimated by inverting parallaxes. $M_{\text{cvz}}$ is the mass of the convection zone in which photospheric metals are mixed. Stellar masses and cooling ages are interpolated from the Montr\'eal grids \citep{Bedard2020}. Values for $\teff$, $\log{g}$, $\log{(\textrm{H}/\textrm{He})}$ (for all stars), and $\log{(\textrm{Ca}/\textrm{He})}$ (for J2050 only) are adopted from a previous study that used the same stellar model \citep{Coutu2019}.}
\label{tableTargetData}
%\resizebox{\textwidth}{!}{ %If it's slightly too wide
\begin{tabular}{lrrrrrrr}
\hline
SDSS & J0848 & J0939 & J0956 & J1227 & J1313 & J1618 & J2050 \\
\hline
\multicolumn{8}{l}{\textit{Catalogue data}}
\smallskip\\
$\alpha$ (J2000; h\,m\,s) & 08\,48\,57.88 & 09\,39\,42.32 & 09\,56\,45.15 & 12\,27\,33.45 & 13\,13\,37.03 & 16\,18\,01.34 & 20\,50\,59.11 \\
$\delta$ (J2000; $\degr$\,$\arcmin$\,$\arcsec$) & +00\,28\,34.9 & +55\,50\,48.7 & +59\,12\,40.6 & +63\,30\,29.4 & +57\,38\,01.6 & +44\,52\,20.8 & $-$01\,10\,21.9 \\
\textit{Gaia G} (mag) & 18.5 & 16.7 & 18.4 & 18.1 & 16.8 & 19.2 & 19.7 \\
Distance (pc) & $236\pm10$ & $68.6\pm0.3$ & $135\pm2$ & $98\pm1$ & $67.7\pm0.2$ & $253^{+14}_{-12}$ & $263^{+30}_{-25}$ \\
\smallskip\\
\multicolumn{8}{l}{\textit{Stellar parameters}}
\smallskip\\
$\teff$ (K) & $11700\pm400$ &            $9020\pm200$ &           $8720\pm100$ &            $7420\pm80$ &            $8800\pm200$ &            $9870\pm300$ &        $10200\pm300$ \\
$\log{[g\,(\text{cm\,s}^{-2})]}$ & $7.89\pm0.1$ &           $8.07\pm0.05$ &           $8.13\pm0.05$ &           $8.07\pm0.04$ &          $7.99\pm0.06$ &           $7.94\pm0.12$ &           $8.44\pm0.21$ \\
Mass ($M_{\sun}$) & $0.52\pm0.06$ &           $0.62\pm0.03$ &           $0.66\pm0.03$ &           $0.62\pm0.03$ &          $0.57\pm0.04$ &           $0.54\pm0.07$ &           $0.86\pm0.14$ \\
Cooling age (Gyr) &  $0.390\pm0.007$ &  $0.96\pm0.03$ &  $1.20\pm+0.07$ &  $1.60\pm0.07$ &  $0.90\pm0.02$ &  $0.63\pm0.07$ &  $1.3\pm0.3$ \\
$\log{(M_{\text{cvz}}/M_{*})}$ &   $-4.61$ &                 $-4.98$ &                $-5.36$ &                $-5.07$ &                 $-4.68$ &                 $-4.65$ &              $-6.01$ \\
\smallskip\\
\multicolumn{8}{l}{\textit{Sinking timescales} (Myr)}
\smallskip\\
O                &                    8.11 &                    4.16 &                   1.72 &                   3.48 &                    7.45 &                    7.62 &                0.386 \\
Na               &                    6.08 &                    2.98 &                   1.25 &                   2.44 &                    5.31 &                    5.59 &                0.281 \\
Mg               &                    6.20 &                    3.01 &                   1.26 &                   2.45 &                    5.37 &                    5.68 &                0.282 \\
Al               &                    5.60 &                    2.69 &                   1.13 &                   2.18 &                    4.79 &                    5.10 &                0.254 \\
Si               &                    5.75 &                    2.75 &                   1.15 &                   2.21 &                    4.89 &                    5.22 &                0.257 \\
Ca               &                    4.55 &                    2.10 &                  0.882 &                   1.65 &                    3.72 &                    4.06 &                0.198 \\
Sc               &                    3.92 &                    1.80 &                  0.762 &                   1.41 &                    3.18 &                    3.49 &                0.171 \\
Ti               &                    3.70 &                    1.69 &                  0.717 &                   1.32 &                    2.98 &                    3.28 &                0.161 \\
V                &                    3.49 &                    1.59 &                  0.677 &                   1.24 &                    2.80 &                    3.09 &                0.152 \\
Cr               &                    3.52 &                    1.59 &                  0.679 &                   1.24 &                    2.82 &                    3.11 &                0.152 \\
Mn               &                    3.35 &                    1.51 &                  0.644 &                   1.18 &                    2.67 &                    2.96 &                0.144 \\
Fe               &                    3.38 &                    1.52 &                  0.648 &                   1.18 &                    2.69 &                    2.98 &                0.145 \\
Ni               &                    3.34 &                    1.50 &                  0.636 &                   1.16 &                    2.64 &                    2.94 &                0.142 \\
\smallskip\\
\multicolumn{8}{l}{\textit{Photospheric abundances ($\text{log\,Z}/\text{He}$ by number)}}
\smallskip\\
H      &             $<-6.1$\phantom{XXxx} &       $-4.6\pm0.06$ &        $-3.5\pm0.2$ &             $<-4.0$\phantom{XXxx} &        $-4.7\pm0.2$ &             $<-5.9$\phantom{XXxx} &             $<-5.6$\phantom{XXxx} \\
O      &             $<-5.0$\phantom{XXxx} &             $<-5.0$\phantom{XXxx} &             $<-4.2$\phantom{XXxx} &             $<-4.0$\phantom{XXxx} &             $<-5.0$\phantom{XXxx} &             $<-5.0$\phantom{XXxx} &             $<-5.0$\phantom{XXxx} \\
Na     &             $<-5.7$\phantom{XXxx} &             $<-8.0$\phantom{XXxx} &        $-6.5\pm0.2$ &        $-8.3\pm0.2$ &             $<-8.0$\phantom{XXxx} &        $-7.1\pm0.3$ &             $<-7.0$\phantom{XXxx} \\
Mg     &        $-5.9\pm0.2$ &        $-7.0\pm0.2$ &        $-5.2\pm0.1$ &        $-7.3\pm0.2$ &        $-7.9\pm0.2$ &        $-6.3\pm0.2$ &        $-6.2\pm0.2$ \\
Al     &             $<-7.0$\phantom{XXxx} &             $<-7.0$\phantom{XXxx} &        $-6.6\pm0.3$ &             $<-7.0$\phantom{XXxx} &             $<-8.0$\phantom{XXxx} &             $<-7.0$\phantom{XXxx} &             $<-6.5$\phantom{XXxx} \\
Si     &        $-5.8\pm0.3$ &             $<-6.4$\phantom{XXxx} &        $-5.5\pm0.3$ &             $<-7.1$\phantom{XXxx} &             $<-7.0$\phantom{XXxx} &             $<-6.4$\phantom{XXxx} &        $-5.4\pm0.2$ \\
Ca     &        $-7.6\pm0.3$ &        $-8.2\pm0.2$ &        $-7.0\pm0.2$ &        $-8.7\pm0.2$ &        $-9.3\pm0.2$ &        $-7.8\pm0.2$ &        $-7.3\pm0.2$ \\
Sc     &              $<-10$\phantom{XXxx} &              $<-11$\phantom{XXxx} &              $<-10$\phantom{XXxx} &              $<-11$\phantom{XXxx} &              $<-12$\phantom{XXxx} &              $<-11$\phantom{XXxx} &             $<-10$\phantom{XXxx} \\
Ti     &        $-9.2\pm0.2$ &        $-9.8\pm0.2$ &        $-8.9\pm0.2$ &        $-9.7\pm0.3$ &         $-11\pm0.2$ &        $-8.9\pm0.2$ &        $-9.0\pm0.4$ \\
V      &             $<-8.5$\phantom{XXxx} &             $<-9.3$\phantom{XXxx} &             $<-9.0$\phantom{XXxx} &             $<-9.3$\phantom{XXxx} &              $<-10$\phantom{XXxx} &             $<-9.0$\phantom{XXxx} &             $<-8.5$\phantom{XXxx} \\
Cr     &        $-8.1\pm0.3$ &        $-9.0\pm0.2$ &        $-8.2\pm0.3$ &             $<-9.0$\phantom{XXxx} &        $-9.8\pm0.3$ &        $-8.0\pm0.2$ &        $-8.1\pm0.4$ \\
Mn     &             $<-9.0$\phantom{XXxx} &             $<-9.5$\phantom{XXxx} &             $<-9.0$\phantom{XXxx} &             $<-9.0$\phantom{XXxx} &             $<-9.5$\phantom{XXxx} &        $-8.9\pm0.2$ &             $<-8.5$\phantom{XXxx} \\
Fe     &        $-6.1\pm0.2$ &        $-7.2\pm0.2$ &        $-6.4\pm0.2$ &        $-7.5\pm0.2$ &        $-8.1\pm0.3$ &        $-6.4\pm0.3$ &        $-6.5\pm0.3$ \\
Ni     &             $<-7.4$\phantom{XXxx} &        $-8.4\pm0.2$ &        $-7.9\pm0.2$ &        $-8.9\pm0.2$ &             $<-8.8$\phantom{XXxx} &        $-7.5\pm0.3$ &             $<-7.5$\phantom{XXxx} \\
\hline
\end{tabular}
%} %Closing brace for \resizebox, if we need it
\end{center}
\end{table*}

HIRES data were reduced using the \textsc{makee} software package, which outputs heliocentric velocity-corrected spectra shifted to vacuum wavelengths. For these heavily polluted DZ~white dwarf stars, where numerous metal lines are present (especially in blue spectra), it was necessary to rely on an instrumental response correction source to calibrate spectral continuum levels and merge the HIRES echelle orders. No dedicated calibrator was observed for this purpose, and instead one of the DZs found to be mostly featureless (J0805, $G\approx15.5$\,mag) was employed. Polynomial fits to each echelle order were made for J0805 (except around \ion{Ca}{ii}~H and~K) and then used to rectify the instrumental response function for the science targets; slight vertical scaling shifts were further applied to bring overlapping order segments into agreement where necessary. Signal-to-noise ratios in the range 5--18 and 6--26 are estimated in the reduced data near {4100\,\AA} and {6600\,\AA}, respectively.

White dwarf atmosphere models were fitted to the data to determine stellar parameters and elemental abundances, given in Table~\ref{tableTargetData}. The models and methods are described in detail elsewhere, but are summarised below \citep{Bergeron1995, Dufour2007, Dufour2012, Coutu2019}. Cooling ages and masses were interpolated from the Montr\'eal grids\footnote{\href{https://www.astro.umontreal.ca/~bergeron/CoolingModels/}{www.\hspace{0pt}astro.\hspace{0pt}umontreal.ca/\hspace{0pt}$\sim$bergeron/\hspace{0pt}CoolingModels/}} \citep{Bedard2020}, and astrometric data were sourced from the \textit{Gaia}~DR3 catalogue \citep{Gaia2022}.

Values for $\teff$, $\log{g}$, and $\log{(\text{H}/\text{He})}$ were adopted from previous work that fitted the same stellar models as used here to SDSS spectra and photometry \citep{Coutu2019}. Those values were verified by re-fitting using parallaxes from \textit{Gaia}~DR3, photometry from \textit{GALEX} and \textit{WISE}, and including the HIRES spectra when determining $\log{(\text{H}/\text{He})}$. Results were in all cases consistent with published values, though the HIRES spectra were less constraining for $\log{(\text{H}/\text{He})}$.

\begin{figure*}
\includegraphics[width=\textwidth]{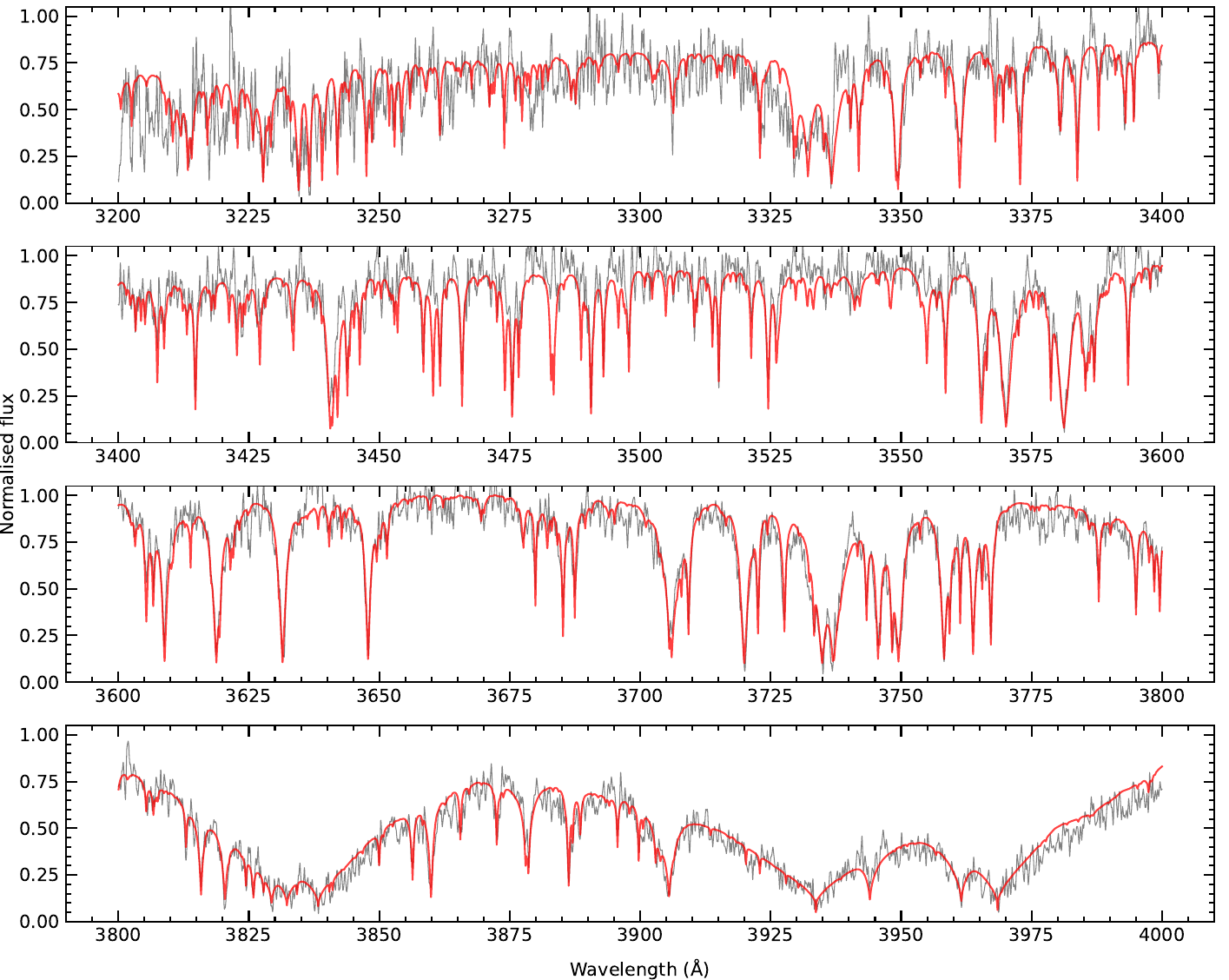}
\caption{HIRES spectrum of J0956 in a region with many metal lines. Smoothing is applied and one cosmic ray hit removed for presentation. The data are shown in grey, while the fitted model is shown in red.} 
\label{figureSpectrum0956}
\end{figure*}

Metal abundances were then determined from the the HIRES spectra, with model atmospheres calculated using atomic data from the Vienna Atomic Line Database \citep{Piskunov1995}. For each element in turn, segments of the data containing lines for that element were then fitted to synthetic spectra across a range of abundances, holding all other elements constant. The process was iterated until convergence. In one case (J2050) a satisfactory fit to the calcium features in the noisy HIRES spectra could not be achieved, so the abundance measured from its SDSS spectrum was adopted \citep{Coutu2019}. To illustrate the data and model fits, Fig.~\ref{figureSpectrum0956} shows a portion of the spectrum of J0956, where many metal lines are visible.

Uncertainties on abundances are challenging to quantify. The location of the spectral continuum is not straightforward to determine, as some regions are blanketed by metal lines. HIRES is an echelle spectrograph, where the data comprise overlapping spectral orders, and the instrumental response may not always be fully corrected. A linear or quadratic correction was therefore included in the fit, to help locate the continuum. Some elements (e.g.~iron) have many lines available in multiple spectral regions, so continuum errors may cancel. However, other elements (e.g.~silicon) can have only one line available, and will be more sensitive to the quality of the continuum fit. For these reasons, the purely statistical abundance uncertainties from the fitting procedure are too optimistic. Uncertainties were therefore estimated from experience of modelling similar stars. Undoubtedly there are also systematic errors due to incomplete or inaccurate atomic line data, or imperfect modelling of the stellar physics \citep{Izquierdo2023}.

\begin{figure}
\includegraphics[width=\columnwidth]{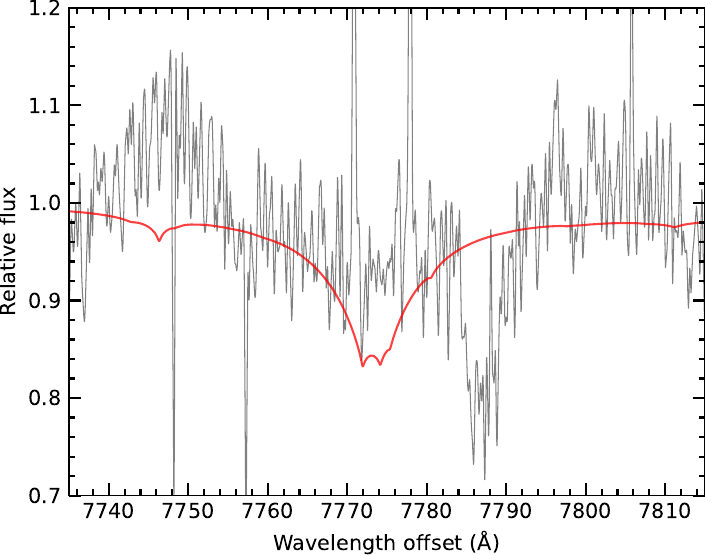}
\caption{HIRES spectrum of J0956 around the \ion{O}{i} triplet near {7775\,\AA}, smoothed and overplotted with a model where the oxygen abundance is set at the upper limit. The narrow features at {7771\,\AA} and {7778\,\AA} are likely sky-subtraction residuals, and the broad feature near {7788\,\AA} is most likely a detector artefact.} 
\label{figureO7775J0956}
\end{figure}

The \ion{O}{i} feature near {7775\,\AA} in the spectrum of J0956 deserves special mention, as the oxygen abundance at that star is important to the discussion in Section~\ref{subsectionJ0956}. The data in that region are noisy, and the continuum normalisation uncertain, so it is challenging to achieve a satisfactory fit to the model. Fig.~\ref{figureO7775J0956} illustrates this, showing the model including oxygen at an abundance equal to the upper limit assigned here. That limit is chosen conservatively, given the poor fit, and the potential influence of an unexpected feature near 7788\,\AA. That feature is likely a detector artefact, as it appears in the spectra of the other stars analysed here, but does not correspond to any atomic line that might be expected to be observed.

\section{Analysis methods}
\label{sectionAnalysis}

Photospheric abundances are now interpreted by comparison with solar system compositions. Previous work has established the utility of this approach (e.g.~\citealt{Xu2013, Hollands2018analysis, Swan2019Xshooter}), and it is updated here to use a Bayesian framework. The models and method are outlined in this section, but a full description including implementation details is given in Appendix~\ref{appendixBayesian}, along with a glossary of the symbols used throughout the paper (Table~\ref{tableSymbols}).

Accretion episodes are modelled by assuming that a mass~$M_{\text{acc}}$ of debris accretes onto an initially-pristine atmosphere at a constant rate for a time~$t_1$, and some further time~$t_2$ may then pass before observation. Photospheric abundance evolution depends on the convection zone mass~$M_{\text{cvz}}$, and the sinking timescale $\tau_k$ for each metal~$k$, which are calculated using a stellar envelope model. Stellar $\teff$ is included as a nuisance parameter, so that its uncertainties are propagated.

Three phases of evolution emerge as $t_1$ and $t_2$ vary \citep{Koester2009}. Photospheric abundances rise together in the \textit{increasing phase}, and their ratios mirror those in the accreted material. As time passes those ratios evolve asymptotically towards fixed values, reaching the \textit{steady state} phase after about five sinking times have passed. When accretion ceases, abundance ratios diverge exponentially as metals drain away in the \textit{decreasing} phase. It is common practice in white dwarf pollution studies to analyse photospheric abundances assuming either that they are in the increasing phase or that accretion has reached a steady state. In a warm, hydrogen-dominated star, sinking timescales are measured in days, and it is therefore reasonable to assume that accretion is ongoing, and has reached a steady state. Neither assumption is safe for DZ~stars, where sinking timescales are on the order of $10^6$\,yr, comparable with estimates of the durations of accretion events \citep{Girven2012, Cunningham2021}. Instead, by inferring $t_1$ and $t_2$ from the data, their (potentially large) uncertainties are included in the analysis.

The photospheric hydrogen abundance before the onset of accretion (expressed as a mass fraction $X_{\text{H},0}$) is also inferred, as it does not sink and thus can only increase with time\footnote{Accretion episodes are assumed to be rapid compared to stellar evolution timescales, so that other sources and sinks of photospheric hydrogen can be ignored \citep{Bedard2023}}. Substantial quantities of hydrogen can be delivered -- for example, if the accreted material is rich in water ice -- but a low observed hydrogen abundance could rule out such a scenario.

The data comprise the observed photospheric abundances~$\bm{X}$, which are compared against photospheric abundances~$\bm{X}_{\text{m}}$ predicted by the accretion model. Many instances of the model are considered, which differ in the elemental mass fractions~$\bm{X}_{\text{a}}$ of the accreted material, drawn from measurements of solar system objects. The majority derive from a meteorite database used in previous work \citep{Nittler2004, Xu2013, Swan2019Xshooter}, where averages are taken of the elemental abundances within each meteorite class, giving higher weight to whole-rock samples. Meteorites are supplemented by several other objects: dust from comet 1P/Halley \citep{Lodders2003}; Earth and Mars (core, primitive mantle, crust, and bulk; \citealt{Wang2018earth, Taylor2008PlanetaryCrusts, Rumble2019CRChandbook, Yoshizaki2020Mars}); and the protosolar composition (metals only; \citealt{Wang2019protosolar}).

The model can accommodate multiple components, specified by their compositions and their fractional contributions. However, introducing more degrees of freedom risks over-fitting the data, so combinations of components should be physically motivated. Mixtures of planetary core, mantle, and crust are considered, based on Earth or Mars templates, to simulate differentiated objects or fragments thereof. Their fractional compositions introduce additional parameters ($f_{\text{core}}, f_{\text{mantle}}, f_{\text{crust}})$, of which only two are independent because $f_{\text{core}}+f_{\text{mantle}}+f_{\text{crust}}=1$. A component of volatile material can also be added, with fraction $f_{\text{volatiles}}$, to account for delivery of ices alongside rocky material.

The data for each star are analysed under each instance of the model, inferring $M_{\text{acc}}$, $t_1$, and $t_2$ in logarithmic form, and using nested sampling to explore the parameter posterior distributions. Priors are set on the parameters to constrain the sample to reasonable solutions, as quantified in Table~\ref{tablePriors} and described below.

\begin{table}
\caption{Priors used in the Bayesian analysis, described further in the text.}
\label{tablePriors}
\begin{center}
\begin{tabular}{llll}
\hline Quantity & Distribution & Min & Max \\
\hline
$P(\log{[M_{\text{acc}}\text{\,(g)}]})$         & $\sim\mathcal{N}(18,3)$    & 9 & 30 \\
$P(\log{[t_1\text{\,(yr)}]})$       & $\sim\mathcal{N}(6.1,1.4)$ & 0 & 10 \\
$P(\log{[t_2\text{\,(yr)}]})$       & $\sim\mathcal{N}(0,3)$     & 0 & $\log{(10\tau_{\text{Ca}})}$ \\
$P(f_{\text{core}}, f_{\text{mantle}}, f_{\text{crust}})$ & $\sim\text{Dirichlet}(\bm{\alpha}=1)$ & -- & -- \\
$P(\teff\textrm{\,[K]})$ & $\sim\mathcal{N}(\teff, \sigma_{\teff})^{\text{*}}$ & 0 & -- \\
$P(\log{X_{\text{H},0}})$ & $\sim\mathcal{N}(-4,4)$ & $-12$ & 0 \\
\hline
\end{tabular}
\end{center}
\flushleft
\footnotesize{$^{\rm *}$ $\teff$ and $\sigma_{\teff}$ refer to the values in Table~\ref{tableTargetData}.}
\end{table}

A Gaussian prior is placed on stellar $\teff$, whose mean and standard deviation are, respectively, the value and uncertainty adopted in Section~\ref{sectionObservationsAndModelling}.  A weakly informative prior is set on $\log{M}_{\text{acc}}$ to favour parent bodies with masses within a few orders of magnitude of large asteroids. Combining photospheric metal masses at cool, helium-atmosphere stars with instantaneous accretion rates at warm, hydrogen atmosphere stars yields an estimate for accretion event timescales, which is used as the prior on $\log{t_1}$ \citep{Cunningham2021}. A wide Gaussian prior is set on $\log{t_2}$. It is truncated at a lower limit sufficiently small to permit values of $t_2\ll \tau_{\text{Ca}}$ that are indistinguishable from ongoing accretion ($t_2=0$), and at an upper limit far into the decreasing phase, where accretion of major planets would be required for metals to remain detectable.

For the core--mantle--crust mixtures, a Dirichlet distribution with concentration parameter $\bm{\alpha}=(1,1,1)$ is used for the prior on $f_{\text{core}}$, $f_{\text{mantle}}$, and $f_{\text{crust}}$, corresponding to a uniform distribution for each between 0 and 1 while enforcing summation to unity. The prior reflects the diversity of material seen at polluted white dwarfs, and the lack of constraints on its origins. Arguments can reasonably be made for sources of accreted material that could generate values at different extremes: surface fragments of an impact-stripped terrestrial planet would exhibit a high $f_{\text{crust}}$, while interior fragments of a differentiated planetesimal disrupted in a hit-and-run collision would exhibit a high $f_{\text{core}}$ \citep{Zuckerman2011,Asphaug2006}. A flat prior lets the data speak for themselves.

\begin{figure*}
\includegraphics[width=\textwidth]{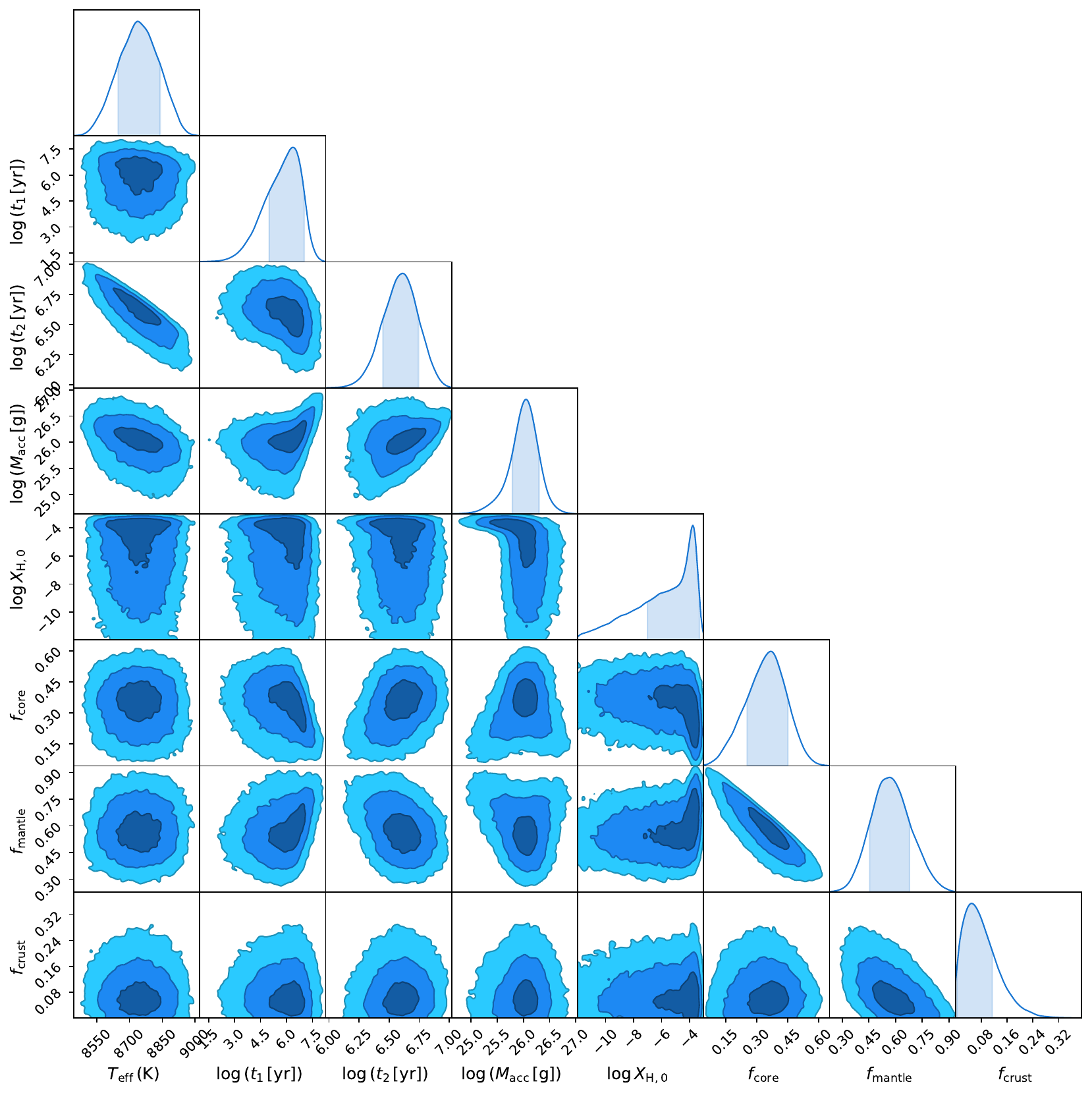}
\caption{Posterior distributions for J0956, inferred under the core--mantle--crust model, using Mars abundances. Contours represent 1-, 2-, and 3-$\sigma$ boundaries.}
\label{figureCornerPlot0956}
\end{figure*}

As an illustrative example of the method, Fig.~\ref{figureCornerPlot0956} shows the posterior distribution for J0956 under a mixture model based on Mars abundances. The marginal distributions of most parameters are approximately Gaussian, but a few features deserve mention. A correlation is seen between $\teff$ and $\log{t_2}$, reflecting the temperature dependence of sinking timescales. For $t_1\gtrsim10^6$\,yr, a correlation with $M_{\text{acc}}$ develops, as accretion moves from the increasing phase towards a steady state and accreted metals begin to sink out of sight. Similarly, as $f_{\text{core}}$ increases, $X_{\text{H,0}}$ decreases, as the hydrogen abundance of core Mars is higher than that of the mantle, and sufficient to supply a non-negligible fraction of the observed photospheric hydrogen. There is a correlation between $\log{t_2}$ and $M_{\text{acc}}$ because the further the star is observed into the decreasing phase, the higher the original mass must have been to achieve the present levels of pollution as metals continue to sink away.

Nested sampling facilitates comparison between models. The method is described in Appendix~\ref{appendixBayesian}, but its salient features are that the relative performance of two models $\mathcal{M}_{\text{a}}$ and $\mathcal{M}_{\text{b}}$ can be measured by their posterior odds $O_{\text{ab}}$, and that model complexity is penalised. A value of $\ln{O_{\text{ab}}}\gtrsim5$ is treated as strong evidence that $\mathcal{M}_{\text{a}}$ is superior to $\mathcal{M}_{\text{b}}$. Further, posterior distributions from two or more models can be averaged, weighted by their posterior odds, effectively marginalising over the models to obtain a model-independent estimate of the parameters. These techniques are central to this study, as the actual composition of the material being accreted cannot be measured directly.

\section{Results}
\label{sectionResults}

\begin{figure*}
 \vspace{0pt}
 \includegraphics[width=\columnwidth, valign=t]{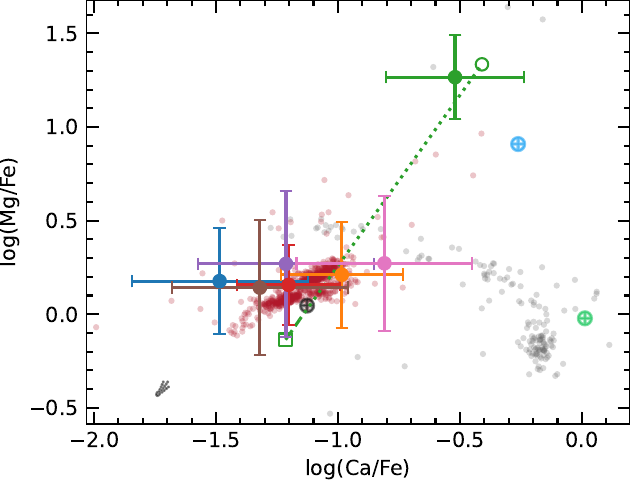}\hfill
 \includegraphics[width=0.85\columnwidth, valign=t]{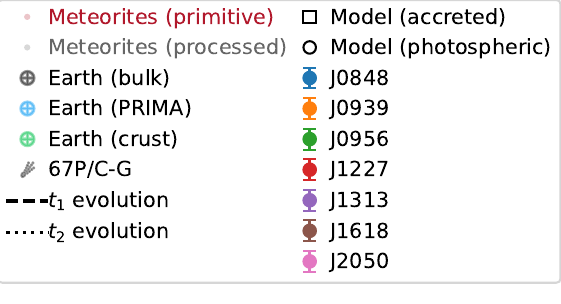}
 \includegraphics[width=\columnwidth]{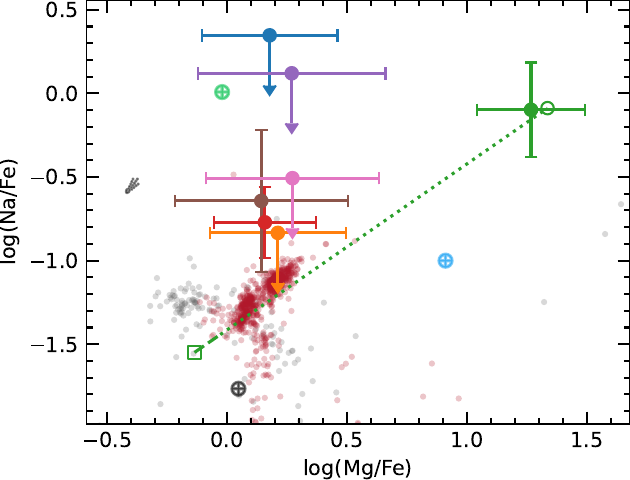}
 \hfill
 \includegraphics[width=\columnwidth]{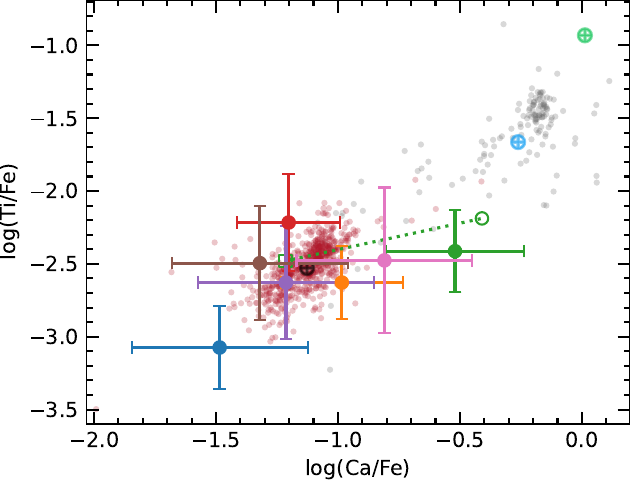}
 \caption{Photospheric abundance ratios and their uncertainties. Arrows denote upper limits. In the case of J0956, predictions of accreted and photospheric ratios are shown for the Mars core--mantle--crust model. The evolution of abundance ratios during and after accretion is shown by dashed and dotted lines, respectively (the dashed segments are much shorter). Solar system compositions are shown for comparison, with meteorites coloured to distinguish primitive (including primitive achondrites) from processed material.}
 \label{figureElementRatios}
\end{figure*}

Before examining the outcome of model comparison and averaging, interesting results emerge simply from inspection of plots of abundances. Fig.~\ref{figureElementRatios} shows some examples, where ratios for comparison objects are shown alongside photospheric ratios for the targets. Each panel tells the same story: in most cases, the photospheric abundances closely correspond to those measured in meteorites with primitive compositions, and are clearly distinct from processed material. There is an outlier, namely J0956, whose photospheric abundances do not correspond to any solar system object. The reason is that differential sinking of metals in the decreasing phase has caused abundances to diverge significantly from those in the accreted material. Fig.~\ref{figureElementRatios} demonstrates this by showing the evolution of the best-fit model abundances from onset of accretion to observation. As metals drain out of the convection zone during the decreasing phase, the observed metals represent only a small fraction of the total accreted mass, which was similar to that of the Moon.

While inspection of abundance-ratio plots can give immediate, heuristic results, the Bayesian framework allows a quantitative assessment of bulk chemistry. Posterior odds are now used to judge how well the models perform relative to each other, so that the best-fitting composition(s) can be identified. Those models are then validated by comparing their predicted abundances against the data, to confirm that a good fit is indeed achieved. Finally, the posterior distributions are averaged together for each star, giving model-independent estimates for the inferred parameters.

\begin{figure*}
\includegraphics[height=15cm]{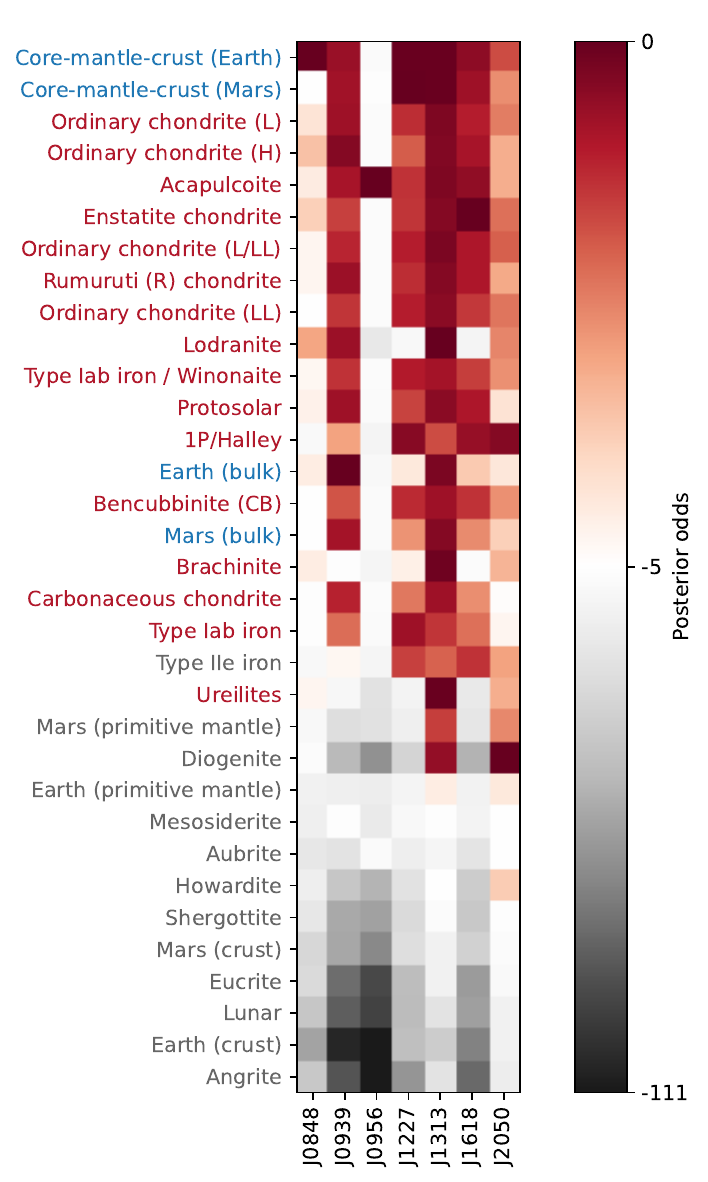}
\hfill
\includegraphics[height=15cm]{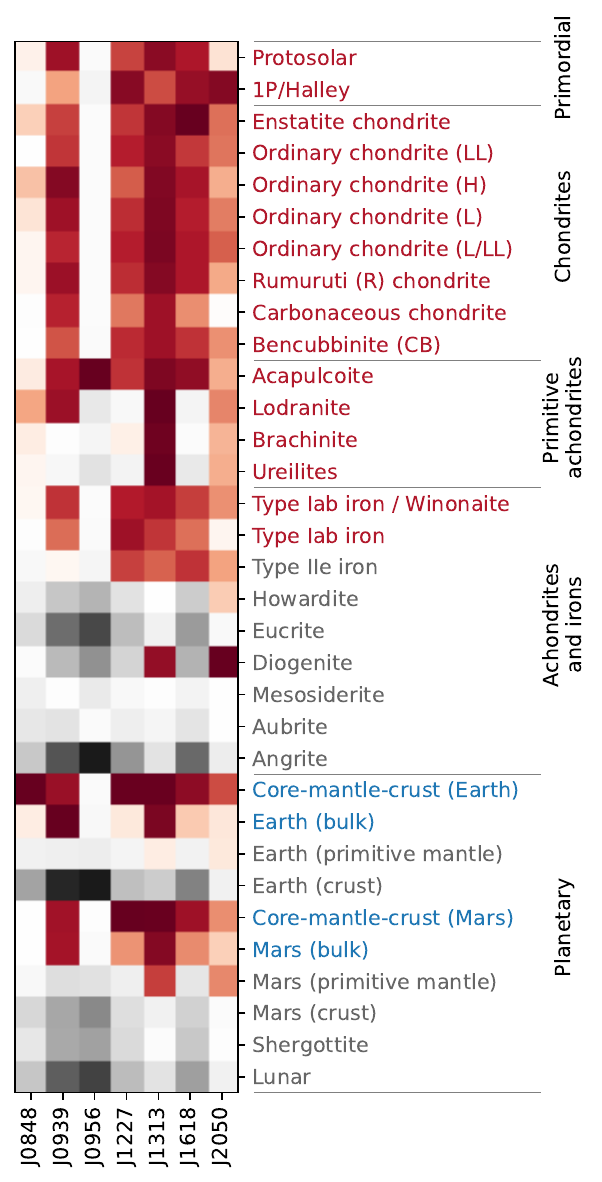}
\caption{Relative performance of solar system compositions and core--mantle--crust mixtures in reproducing photospheric abundances. The posterior odds relative to the best model at each star are indicated in colour, where the best are deep red, fading to white as models approach the $\ln{O}<-5$ threshold for being strongly disfavoured, and then to shades of grey beyond that. Labels on the $y$-axis are coloured red for primitive compositions (including primitive achondrites), grey for processed material, and blue for planet-based models. In the left-hand panel, the models are ordered by their average rank across all stars, while in the right-hand panel they are ordered approximately by degree of processing.}
\label{figureComparisonEvidence}
\end{figure*}

Fig.~\ref{figureComparisonEvidence} compares the posterior odds (Equation~\ref{equationPosteriorOdds}) among all the compositional models, relative to the best model for each star. The best model is coloured dark red, with other models assigned increasingly lighter shades as their relative odds become less favourable. Models with $\ln{O}<-5$ are strongly disfavoured compared to the best model, and are represented in shades of grey. To aid interpretation, three categories of model (primitive, modified, and planetary) are used to assign label colours, and models are ordered by their overall performance. Specifically, the models for each star are ranked by their odds, and the average rank of each model across the sample determines the ordering.

Among the solar system object models, primitive materials such as chondritic meteorites perform well, while more processed compositions are strongly disfavoured. Chondrites and primitive achondrites (e.g.~acapulcoite) are little modified since their formation early in solar system, and are highly ranked for most stars in the sample. Other achondrites, which score lower, are likely fragments of objects that have been modified since their formation. For example, howardite, eucrite, and diogenite (HED) meteorites are each thought to originate from different localised sub-regions of 2~Vesta, a differentiated minor planet, and are decisively ruled out at most of the stars. The model comparison outcome therefore suggests that the target stars are accreting primitive material.

One model -- acapulcoite -- is strongly favoured over all others considered for J0956, whereas several models are possible for the rest of the sample. Rather than the star accreting material that is entirely different to that at the other stars, it is more likely that the models are simply better distinguished from each other owing to the number of metals detected.

\begin{figure}
\includegraphics[width=\columnwidth]{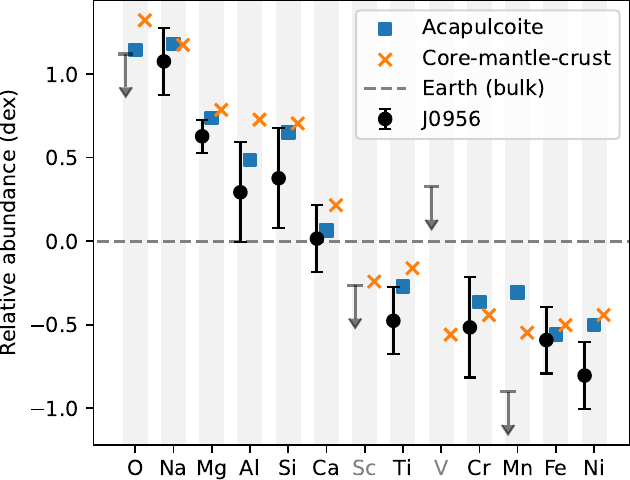}
\caption{Example of model fits at J0956. Photospheric abundances are shown in black, with downward arrows indicating upper limits. Coloured symbols show projected photospheric abundances using median posterior parameters for acapulcoite and the core--mantle--crust mixture based on Mars. All abundances are normalised to those of bulk Earth, represented by the dotted line, and the mean of the photospheric abundances defines the arbitrary zero point. Elements labelled in grey are not used in model fitting as their abundances are not available for all solar system compositions.}
\label{figureModelPredictions0956}
\end{figure}

While the comparison exercise establishes rankings for the models, it does not indicate whether any are reasonable, so a comparison against the data is also required. Goodness-of-fit measures such as reduced-$\chi^2$ require well-determined abundance uncertainties, which are difficult to achieve for white dwarfs. However, visual inspection confirms that the top-ranked models predict photospheric abundances that are in reasonable  agreement with the data. An example is shown in Fig.~\ref{figureModelPredictions0956}, which compares the data against the best models from each family for J0956, calculated using median posterior values for each parameter. The upper limit for manganese is potentially in tension with model predictions, but otherwise the data are a good fit to the models. The other target stars show similar agreement with their best-fitting models, and thus it is reasonable to interpret the results in the context of those models.

Core--mantle--crust mixture models are competitive with the chondritic composition models, despite their additional complexity. By contrast, the single-component models based on bulk planetary compositions are not highly favoured, though in most cases they cannot be completely ruled out. The fact that the mixture models outperform the bulk compositions suggests that their variable proportions are the key to their success. Only at J0956 are the mixture models rejected: the one based on the Mars template is ranked in second place, but its posterior odds versus the leading model are unfavourable.

\begin{figure}
 \includegraphics[width=\columnwidth]{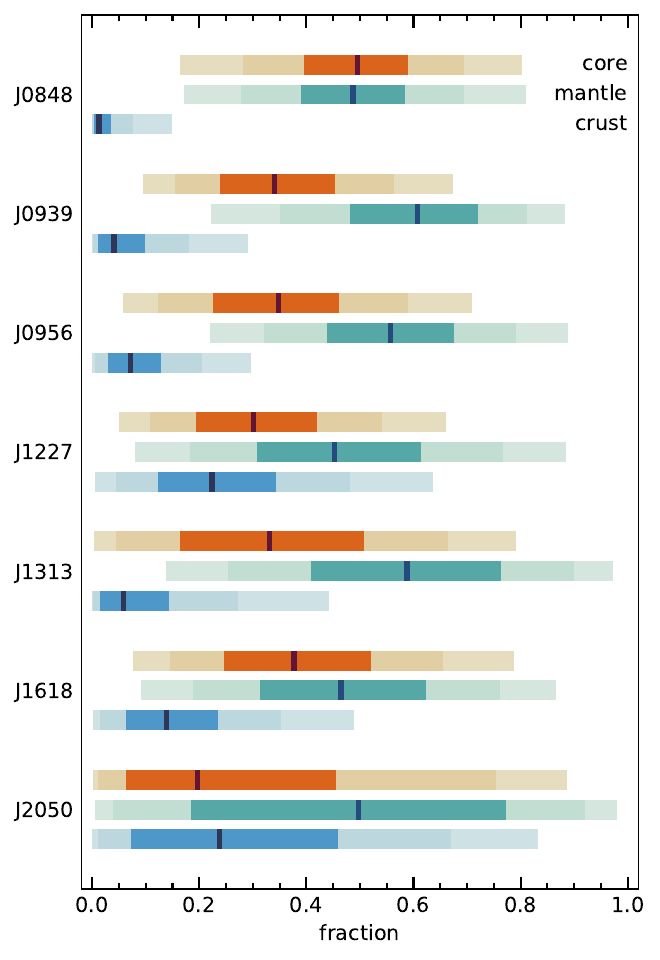}
 \caption{Core, mantle, and crust fractions inferred under the mixture models, averaged over both sets of abundances (Earth and Mars). Coloured bars indicate the 1-, 2-, and 3-$\upsigma$ credible intervals of the marginalised distributions, with a dark mark at the median value.}
 \label{figureCoreMantleCrust}
\end{figure}

Evidence-weighted average posterior distributions for the core, mantle, and crust fractions are shown in Fig.~\ref{figureCoreMantleCrust}. Note that only $f_{\text{core}}$ and $f_{\text{crust}}$ are actually sampled, as $f_{\text{mantle}}$ is fully determined by the other two. The marginal distributions do not capture the correlations that are present, particularly between $f_{\text{core}}$ and $f_{\text{mantle}}$, but examples can be seen in Fig.~\ref{figureCornerPlot0956}. In general, a large spread of compositions are possible, but the same pattern repeats across the sample: accreted material typically appears mantle-rich, with a substantial core fraction, and a small crust fraction.

As the best-performing models provide reasonable fits to the data, meaningful estimates of the parameters of their ongoing or most recent accretion events can be made from evidence-weighted averages, which are given in Table~\ref{tableAverage}. Parent body masses are inferred to be in the range $10^{22}$--$10^{26}$\,g, consistent with minor planets approximately $10^2$--$10^3$\,km in radius. Accretion has lasted for $10^5$--$10^7$\,yr, comparable to the sinking timescales, but with broad enough uncertainties that accretion could be (or have been) anywhere between the increasing phase and a steady state. The similarity of this range of times to the prior on $t_1$ means that no strong conclusions can be drawn about the duration of accretion events from this dataset.

For six of the stars, accretion is inferred to have ceased only $t_2\sim10^2$\,yr ago, orders of magnitude less than their sinking timescales. Such values are merely artefacts of sampling in logarithmic space, and are essentially equivalent to zero, indicating that accretion is likely ongoing. Only J0956 appears to have ceased accreting, around 4\,Myr (four sinking timescales) ago, so that photospheric metals have since been substantially depleted. Photospheric abundance ratios have diverged exponentially (see Equation~\ref{equationAccretionT2}), and thus appear as outliers in Fig.~\ref{figureElementRatios}. Likewise, a trend towards higher abundances of lighter elements (which have longer sinking timescales) is evident in the photospheric metals, as shown in Fig.~\ref{figureModelPredictions0956}. At J2050, the posterior for the Earth mixture model is bimodal, admitting both ongoing accretion of a mantle-rich mixture, or accretion of a core-rich mixture that ceased up to two sinking times ago. However, the best-performing solar system composition models all favour ongoing accretion. On balance, therefore, ongoing accretion is favoured at J2050, but additional abundances (e.g.for sodium and nickel) are needed to confirm that.

\begin{table*}
\caption{Estimates of accreted mass and episode duration, derived by evidence-weighted averaging over all models. Negligibly small values for $t_2/\tau$ are reported as zero, as they represent ongoing accretion.}
\label{tableAverage}
\begin{center}
\begin{tabular}{lllllll}
\hline
Star & $\log{[M_{\text{acc}}\,(\text{g})]}$ & $\log{[\dot{M}_{\text{acc}}\,(\text{g\,s}^{-1})]}$ & $\log{[t_1\,(\text{yr})]}$ & $\log{[t_2\,(\text{yr})]}$ & $t_1/\tau$  & $t_2/\tau$\\
\hline
\rule{0pt}{1.5em}J0848 &  $24.4^{+0.5}_{-0.5}$         &  $9.8^{+0.7}_{-0.1}$ &  $7.1^{+0.5}_{-1.0}$ &    $2.4^{+3.1}_{-1.6}$ &  0.4--10 &  0--0.1 \\
\rule{0pt}{1.5em}J0939 &  $22.9^{+0.8}_{-0.3}$ &   $9.1^{+1.2}_{-0.2}$ &    $6.3^{+1.0}_{-1.4}$ &    $2.0^{+2.1}_{-1.4}$ &  0.04--10 &  0 \\
\rule{0pt}{1.5em}J0956 &  $26.0^{+0.3}_{-0.4}$ &   $12.5^{+1.2}_{-0.7}$ &   $5.9^{+0.9}_{-1.2}$ &  $6.59^{+0.14}_{-0.15}$ &  0.05--8 &  3.8--5.5 \\
\rule{0pt}{1.5em}J1227 &  $22.6^{+0.8}_{-0.3}$ &   $8.7^{+1.0}_{-0.2}$ &   $6.4^{+0.9}_{-1.3}$ &    $2.0^{+2.0}_{-1.3}$ &  0.08--10 &  0 \\
\rule{0pt}{1.5em}J1313 &  $22.1^{+0.6}_{-0.3}$ &   $8.4^{+1.3}_{-0.5}$ &    $6.1^{+1.2}_{-1.4}$ &    $2.0^{+2.1}_{-1.4}$ &  0.01--5 &  0 \\
\rule{0pt}{1.5em}J1618 &  $23.9^{+0.6}_{-0.4}$ &   $9.8^{+1.3}_{-0.3}$ &   $6.4^{+0.9}_{-1.4}$ &    $2.0^{+2.2}_{-1.4}$ &  0.03--7 &  0 \\
\rule{0pt}{1.5em}J2050 &  $23.8^{+0.9}_{-0.7}$ &   $10.2^{+1.3}_{-0.2}$ &  $5.9^{+0.8}_{-0.9}$ &    $2.8^{+2.8}_{-2.0}$ &  0.4--20 &  0--2 \\
\hline
\end{tabular}
\end{center}
\end{table*}

The results obtained above do not take into account the upper limits for scandium or vanadium, as measurements of those elements are not present in all of the solar system compositions, as required for model averaging and comparison. For completeness, parameter inference is repeated with scandium and vanadium included in the data, and it is verified that they have a minimal influence on the outcomes for the best-performing models at each star.

\section{Discussion}
\label{sectionDiscussion}

The key findings of the previous section are summarised as follows, and will be discussed below:

\begin{itemize}
  \item J0956 is observed in the decreasing phase, several Myr after accretion ceased, so that its photospheric abundances have evolved substantially from those of the accreted parent body.
  \item The target stars are accreting material with compositions similar to primitive solar system objects such as chondrites.
  \item Core--mantle--crust mixture models perform well, providing a good fit at six out of the seven stars, with no clear preference for Earth or Mars as a compositional template.
  \item Accretion episodes have lasted around $10^5$--$10^7$\,yr, comparable to sinking timescales at these stars, so accretion has likely moved beyond the increasing phase, and photospheric metal abundances do not exactly reflect those of the accreted material. However, this result is influenced by a prior, and should not be treated as an independent measure of accretion event lifetimes.
  \item Inferred accreted masses correspond to large asteroids or dwarf planets ($M_{\text{acc}}\ge10^{22}$\,g) at most of the stars, and an object with a mass similar to that of the Moon ($10^{26}$\,g) at J0956.
\end{itemize}

The stars analysed here are representative of the DZ~spectral class, so these findings should be broadly applicable to the wider population. There is one potential exception, in that J2050 has a higher mass than typical white dwarfs and the rest of the sample, corresponding to a late B-type main sequence progenitor with a mass of $3.6^{+1.3}_{-0.8}$\,$M_{\sun}$ \citep{Cummings2018}. It thus joins the small but growing population of evolved planetary systems that formed around stars larger than 3\,$M_{\sun}$ \citep{Coutu2019, Hollands2021alkali}. Despite the apparently lower efficiency of rocky planet formation in such systems \citep{Lisse2019}, the bodies that do form appear to have similar chemistry to those polluting most white dwarfs, whose progenitors were 1--2\,$M_{\sun}$ stars.

\subsection{The planetary body polluting J0956}
\label{subsectionJ0956}

The most polluted star in the sample -- J0956 -- also has the most metals (nine) identified in its spectrum. Two recent analyses of this white dwarf have offered different interpretations of the composition of the accreted material. A population study concluded that it is core-like, and observed in the decreasing phase \citep{Harrison2021bayesian}. A more focused study based on fresh data concluded that it is rich in magnesium silicates and water ice (\citealt{Hollands2022coolWhiteDwarfs}, hereafter H22). In contrast to those inferences, this paper presents a third interpretation: accretion of dry, primitive material, observed far into the decreasing phase. The $t_2$ parameter is tightly constrained to be in the decreasing phase, where accretion ceased around 4\,Myr ago. Since then, downward diffusion of metals has diminished their photospheric abundances by almost two orders of magnitude, yet it remains heavily polluted. The parent body had a mass of around $10^{26}$\,g, comparable to the largest solar system moons, and perhaps even larger, as half of its mass may have been ejected from the system during tidal disruption \citep{Malamud2020A}. Such an object would almost certainly have been differentiated. The primitive composition indicated by the analysis here is consistent with the entire object being accreted in a single event.

A tentative detection of photospheric carbon has been reported at J0956, based on ultraviolet spectral features (H22). That study favoured dredge-up from the stellar core rather than accretion as the origin of the carbon, owing to its high (albeit poorly-determined) abundance. An alternative explanation for the high abundance of carbon is that its long sinking timescale relative to other metals causes it to become concentrated during the decreasing phase of accretion. That scenario was rejected by H22 as it requires accretion of a large object to achieve the observed abundances, and such objects are rare. However, that statistical argument is weakened because J0956 was not targeted at random: it is one of the most heavily-polluted stars yet identified, so the object it accreted may also be an outlier.

An over-abundance of oxygen compared to other rock-forming elements may indicate accretion of volatile-rich material \citep{Farihi2013}, but can also be caused by concentration in the decreasing phase. Only an upper limit can be determined from the data analysed here, but oxygen features are present in the optical spectra of J0956 analysed by H22. Equivalent widths varied between instruments without an obvious explanation, so abundances were reported separately for each dataset. Both measurements suggest there is more oxygen than could be accounted for by ongoing accretion of dry rock. Given that the most recent accretion event appears to have concluded and is now observed in the decreasing phase, it is important to assess whether any oxygen was delivered by accretion of volatiles during that event.

The compositional mixture model is extended to include volatiles, with fraction $f_{\text{volatiles}}$, and the flat Dirichlet prior is retained. Two compositions are considered for the volatile component: pure water ice, and the comet 67P/Churyumov--Gerasimenko (67P/C--G; \citealt{Rubin2019}). The abundance measurements and limits found here for J0956 are analysed under the models that include volatiles. All are decisively rejected compared to the best-performing volatile-free models, and thus a volatile component is not supported by the data.

The analysis is repeated using stellar parameters, abundance measurements, and sinking timescales from H22, for both the original models (without volatiles) and their extensions (with volatiles). Four combinations of abundances are investigated: using the oxygen abundances measured from their GTC or SDSS spectra ($\log{\textrm{O}/\textrm{He}}=-4.6\pm0.1$ and $-4.1\pm0.2$, respectively), and including or excluding a carbon abundance of $\log{(\textrm{C}/\textrm{He})}=-4.65\pm0.55$. Posterior odds are calculated by treating the models including volatiles as separate model families.

\begin{figure}
\includegraphics[width=\columnwidth]{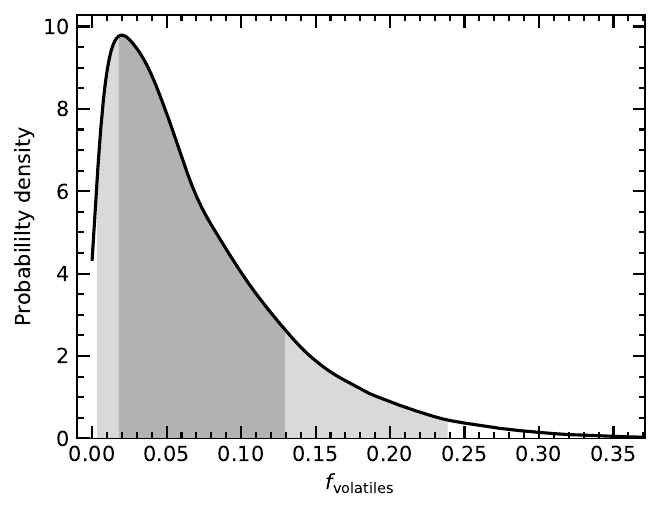}
\caption{Posterior for the volatile fraction at J0956, averaged across all models that include that parameter (the best-performing model does not). Abundances from H22 are analysed, using the oxygen measurement from their GTC data, and excluding the tentative carbon measurement. \textit{Dry rock is strongly favoured for all other combinations of data}, so posteriors for those cases are not plotted. Shaded regions indicate the 1- and 2-$\upsigma$ credible intervals.}
\label{figureJ0956volatiles}
\end{figure}

Volatile-bearing composition models are strongly disfavoured for the H22 data, except where the oxygen abundance measured from the GTC spectrum is used, and carbon is excluded. Even in that case, the best models are dry, namely a type~Iab iron meteorite or a Mars mixture model. However, a few volatile-enriched models (both primitive meteorites and planet mixtures) also return posterior odds above the threshold of $\ln{O}>-5$. Their averaged posterior for $f_{\text{volatiles}}$ is strongly weighted towards low values, with a median of 6~per~cent and a 1-$\upsigma$ upper bound of 13~per~cent (Fig.~\ref{figureJ0956volatiles}). It may seem counterintuitive that using the lower oxygen abundance from GTC data can admit volatile-enriched models, but when the SDSS measurement is used, higher values of $t_2$ are preferred. In other words, the higher photospheric oxygen abundance is more readily explained by observation in the decreasing phase than by the accretion of ices. Overall, there is minimal evidence for a substantial volatile component in the accreted material at J0956, either for the abundances determined either here or by H22 (which are consistent at the 2-$\upsigma$ level for all elements common to both studies).

\subsection{Compositions}
\label{subsectionCompositions}

The rocky material accreted by all seven stars is similar to that found at the majority of polluted white dwarfs analysed in detail so far. There appear to be no dramatic outliers among the parent body compositions, although the major rock-forming elements (O, Mg, Si, Fe) are not all detected simultaneously at any of the targets, so there is still room for surprises. There are hints that J1227 and J1618 have elevated sodium abundances compared to solar system objects, potentially indicating the parent objects formed further out in their protoplanetary disks, but the uncertainties are too large to draw a definitive conclusion.

Comparison against solar system objects finds that primitive compositions provide a good match for the exoplanetary material observed across the sample, in common with many other polluted stars \citep{ Xu2019composition, Hoskin2020, Izquierdo2021GD424, Harrison2021bayesian}. The results here are comparable with recent work assessing the similarity between white dwarf pollution and CI chondrites \citep{Trierweiler2023}. That study considered 31 white dwarfs polluted with oxygen and several other elements, and found that over half were accreting material consistent with chondrites. Those stars are almost all hotter than the ones analysed here, for which chondritic compositions are also favoured in the majority of cases. Taken together, these complementary samples show a broad consistency in the compositions of accreted objects across a wide range of cooling ages.

Chondritic abundances can be interpreted in more than one way. They could reflect accretion of chondritic fragments, little modified since their formation, but they could also result from accretion of large objects, whose averaged compositions are primitive, obscuring any differentiation or other internal processing. The smallest accreted mass among the stars analysed here is $10^{22}$\,g at J1313, comparable to a large member of the solar system asteroid belt. If formed under similar conditions to those in the early solar system, such an object is expected to have differentiated through radiogenic heating \citep{Moskovitz2011, Monnereau2023}. The objects sampled here therefore appear not to be the products of catastrophic collisions that separated the layers of larger parent bodies sufficiently to result in fragments with extreme core- or mantle-rich compositions. Tidal disruptions at the Roche radius can also separate the layers of a differentiated object so that the composition of material being accreted changes with time \citep{Brouwers2022asynchronous}. However, the sinking timescales of the stars in this sample are longer than the expected timescale of any such chemically-fractionated accretion. Cool, helium-atmosphere stars are therefore not ideal targets with which to probe differentiated objects, as in most scenarios photospheric abundances are likely to resemble the primitive compositions found here, as parent-body differentiation or fractionated accretion are averaged out.

The core--mantle--crust mixture models perform well, ranking highly at six out of seven stars, but there is no clear consensus from the results whether Earth- or Mars-like abundances better reflect those in other planetary systems. The only object where a mixture of Earth abundances were clearly superior to Mars abundance is J0848, which has the largest accreted mass among the stars where a mixture model is favoured. Mars has a shorter and simpler formation history than Earth, and is expected to be representative of a planetary embryo \citep{Yoshizaki2021Mars}. It is tempting to interpret the relative performances of the mixture models in this light, and to argue that J0848 must have had a formation history more similar to that of Earth than of Mars, but there is no other evidence to support such a claim. Indeed, the mixture model indicates a larger core fraction for J0848 than the other stars, and a crust fraction tightly constrained to be small, pointing towards a composition more like that of Mercury than of Earth or Mars. The composition of Mercury is not yet known in sufficient detail to be used as a template for the analysis performed here \citep{Nittler2018}, so no claims are made in this regard. However, if an oxygen abundance can be determined at J0848, that would permit an analysis of the oxygen fugacity of the accreted material, with the potential to reveal whether it formed in reducing conditions, like Mercury \citep{Doyle2020}.

Nine elements are detected at J0956, which proves sufficient to resolve models from each other such that acapulcoite meteorites are selected as the best match to the accreted material. Interestingly, acapulcoites are thought to share the same parent body as lodranites; the two meteorite types are very similar except that lodranites have experienced greater loss of partial melt than acapulcoites, probably due to formation deeper in the parent body \citep{Weisberg2006}. This means lodranites are depleted in plagioclase and troilite relative to acapulcoites, which are themselves depleted in these species relative to chondrites. The body accreted by J0956 is orders of magnitude more massive than the parent body of acapulcoites, and it does not seem possible it could have accreted only material from a particular depth on an acapulcoite-like parent body. Instead, the following conjecture is offered: J0956 may have accreted mantle material from a large body that had lost iron to the core, and a plagioclase component to the crust, which was likely lost at an earlier time; the lost amounts of troilite and plagioclase happen to match acapulcoites.

Spectra of sufficient quality to detect around nine or more metals appear necessary for detailed analyses, such as that just presented, where meteorite classes must be clearly distinguished from each other. With only a few elements detected at most of the stars in the sample, however, there is a limit to the information that can be gleaned from their abundances. For example, two recent studies offer interesting ways to probe processes operating during planet formation, but they are of limited utility for the data presented here. The first presents a method for inferring parent body mass independently of photospheric metal mass \citep{Buchan2022}. Comparison against fig.~6 in that work shows the abundance ratios measured in this sample fall within the region of parameter space where the geochemical model has the least predictive power. The second study proposes that the object accreted by GD\,362, a heavily-polluted helium-atmosphere star, provides evidence for post-nebular volatilisation \citep{Harrison2021postNebula}. In that scenario, heat-driven loss of volatile elements from planetesimals after their formation can be detected from sodium, magnesium, and manganese abundances. Those abundances are measured here for J1618, and limits are determined for other stars in the sample. Their ratios provide no evidence for post-nebular volatilisation at any of the stars considered, although the limits are weakly constraining.

Comparison against solar system objects provides an empirical way to assess white dwarf pollution. This both complements and contrasts with recent work, where photospheric abundances are predicted using a complex theoretical model that considers the full history of accreted objects from formation to destruction \citep{Harrison2021bayesian}. Each approach has strengths and weaknesses. The empirical method has the advantage of simplicity, but also benefits from comparison objects that can be analysed in the laboratory, whose parent bodies have good prospects of being increasingly well understood as research continues. However, that is also a potential weakness of the method: it relies on planetesimals accreted by white dwarfs being sufficiently similar to solar system objects that insights from one can be transferred to the other. There are promising indications that this is generally the case \citep{Doyle2019}, but any exotic outliers will resist an empirical comparison. Such objects may remain amenable to theoretical modelling, whose strength lies in its flexibility. At present, such models are over-specified compared to the available data and associated uncertainties, and suffer strong degeneracies \citep{Buchan2022}. However, there are good prospects for constraining those models if the composition of protoplanetary disks can be inferred from resolved companions that are still on the main sequence \citep{Bonsor2021}.

\subsection{Accretion histories}
\label{subsectionAccretion}

Time-averaged accretion rates $\dot{M}_{\text{acc}}$ in the range $10^{8}$--$10^{10}$\,g\,$\text{s}^{-1}$ are inferred for six of the stars, and are typical of the wider population, but the rate at J0956 was at least $10^{12}$\,g\,$\text{s}^{-1}$, and thus likely to have been detectable in X-rays while ongoing \citep{Farihi2018Magnetism, Cunningham2022Xrays}. In this simple model, $\dot{M}_{\text{acc}}$ is assumed to be constant, but in reality that cannot be guaranteed. A sharp increase in $\dot{M}_{\text{acc}}$ that delivers sufficient material to dominate the photospheric metals will effectively reset accretion to the increasing phase. There is observational evidence that such events occur, and they are also predicted by theory \citep{Farihi2012intenseAccretion, Rafikov2011runaway}. However, at present, there is no empirical evidence to guide such a time-varying accretion model.

The decreasing phase cannot be ignored when interpreting metal abundances at white dwarfs. Fig.~\ref{figureModelPredictions0956} illustrates how heavier elements (e.g.~iron, titanium) have become depleted relative to lighter elements (e.g.~sodium, magnesium) at J0956. It is particularly important to consider this possibility where a substantial quantity of oxygen is detected, to avoid the risk of misinterpreting an apparent excess. Only a few of the dozens of stars targeted by 8--10-m class telescopes are likely in the decreasing phase \citep{Swan2019Xshooter, Elms2022}, and a further 12 potential candidates have been identified from over 200 SDSS spectra \citep{Harrison2021bayesian}. These statistics suggest that the observable decreasing phase is at most a few per~cent of the duration of an accretion event. However, those samples are biased towards heavily polluted stars, so their targets are more likely to have been observed while still accreting than in the decreasing phase. Moreover, given the degeneracy between $t_1$ and $t_2$ (Fig.~\ref{figureT1T2degeneracy}), it will be challenging to fully disentangle evolution of abundances towards a steady state from their evolution into the decreasing phase. Reliably characterising accretion events across a population would require knowledge of the mass function of planetesimals, the mean accretion lifetime, the duty cycle of pollution events, and any trends with age in those quantities. Such an investigation might be attempted using a large volume-limited sample, or one whose selection function is well-characterised.

\subsection{Bayesian analysis}
\label{subsectionBayesian}

This study demonstrates the potential power of Bayesian analysis when applied to white dwarf photospheric abundances. However, it is merely a statistical tool and must be used with care, as inappropriate choices of priors or models will generate misleading results. Some benefits and limitations of the analysis employed here are now discussed. It is emphasised that while this study presents quantified results with uncertainties, they are only be as good as the underlying models, whose simplifications and approximations must always be borne in mind.

Monte Carlo sampling allows convenient propagation of uncertainties. In previous work, accretion has often been treated as ongoing, and either in the increasing phase or a steady state, which are strong assumptions, especially at DZ~stars. It is safer to model the accretion episode, so that the unknown accretion history can be marginalised over and built into the uncertainties.

Bayesian methods allow prior knowledge to be incorporated into the analysis. However, judicious choice of priors is essential. They should reflect the knowledge and reasonable expectations held at the start of the experiment, and must not be influenced by inspection of the data. The priors employed here are weakly constraining, compared to the strength of the data, and should not have an undue influence over the results, but future work may be able to improve on them. For example, theoretical models can constrain the photospheric hydrogen abundance expected at a given effective temperature in the absence of accretion \citep{Rolland2018, Bedard2023}. As understanding of accreted compositions improves, some combinations of core, crust, and mantle may be judged unlikely, so the Dirichlet prior could encode those expectations, perhaps in a mass-dependent way \citep{Zuckerman2011}. However, at present there is no clear motivation for why one composition might be favoured over another, and to do so prematurely would undermine the fundamental goal to determine and interpret the observed bulk chemistry of exoplanets.

Once a substantial body of unbiased data is available, there are better prospects for learning and applying a compositional prior to future analyses. The motivation for that would not be to influence results related to bulk chemistry, but to constrain the accretion model parameters. For example, consider a dataset where only calcium, magnesium and iron are detected, and magnesium shows an enhanced abundance. Under a uniform compositional prior (as used here) two potential explanations must be considered: unusual chemistry, or observation in the decreasing phase, where the longer sinking time of magnesium has led it to become concentrated. Should it turn out that material typically accreted by white dwarfs is similar to the bulk Earth, then tuning the Dirichlet prior to favour such a composition would securely identify the decreasing phase as the reason for an enhanced magnesium abundance, and enable a measurement of the time since accretion ceased. Where more than just three elements are detected, it would still be possible to identify outliers with exotic chemistry, as a well-chosen prior should guide results from weak data, but be overwhelmed by strong data.

This study applies many models to the data, which differ in their compositions, but treat accretion in the same way. Nested sampling facilitates both quantitative comparison between the models, and averaging of parameters across them, thus allowing deeper insights than reduced-$\chi^2$ methods \citep{Xu2013, Swan2019Xshooter}. The comparison exercise narrows down the range of likely compositions in the accreted material, but takes into account uncertainties in the accretion history. Meanwhile, averaging over compositional models allows the accretion history to be probed without knowing the exact composition of the accreted material.

Given that the more complex core--mantle--crust mixture models were competitive with the single-composition models, both are reasonable choices for future analyses. The rich library of solar system comparisons available may be better suited to analyses of individual stars where many metals are measured, as the models will be better distinguished from each other in the high-dimensional space (as demonstrated here by J0956). The mixture models may be more useful for population studies, as they require fewer computational resources and are intuitively interpretable.

Turning now to the limitations of this analysis, all metals are treated as coming from a single object, accreted at a constant rate, but photospheric metals may have come from multiple objects, potentially at varying rates \citep{Turner2020, Johnson2023}. Nevertheless, abundances will tend to be dominated by the most recently accreted large object, so the model used here is likely to be a useful approximation.

Sinking timescales depend on atmospheric composition, and therefore vary throughout an accretion episode, but they are held constant in this study. That is a good approximation where accretion is ongoing, but begins to break down in the decreasing phase, as metals drain out of the photosphere. For observable systems, the difference is not dramatic -- at a heavily-polluted star like J0956, a 2-dex decline in metal abundances in the convection zone increases sinking timescales by a factor of a few -- but it means that $t_1$ and $t_2$ will be overestimated.

\subsection{Prospects for future work}
\label{subsectionRecommendations}

The models presented here have up to eight independent parameters, and a similar number of measurements are necessary in order to limit degeneracies. Spectroscopic observations should therefore aim to achieve sufficient S/N and wavelength coverage to measure all of the major rock-forming elements, and ideally several additional trace elements, if complex models are to be employed. It is equally important to report upper limits on abundances, even if they are only weakly constraining, and to use them in subsequent analysis. The oxygen abundance upper limit proved sufficiently informative to rule out the presence of volatiles in the accreted material at J0956.

The model employed here is tailored to helium-atmosphere stars, whose sinking timescales are comparable to the lifetime of an accretion event. However, it can be easily adapted for warm hydrogen-atmosphere stars, where accretion has almost certainly reached a steady state. In that case $\dot{M}_{\text{acc}}$ would replace $M_{\text{acc}}$ and $t_1$. However, inferring $t_2$ may still be useful: if the accretion rate abruptly drops, a short-lived decreasing phase will be observable. Spectroscopic surveys are now covering many thousands of stars, with the potential either to discover examples of such stuttering accretion, or to place constraints on their frequency.

An extension to the model could consider a mixture compositions in a more generic fashion than the $f_{\text{volatiles}}$ parameter used above \citep{Johnson2023}. Such a model would be useful for detecting overlapping accretion events, or a constant background supply of material by small bodies \citep{Wyatt2014,Turner2020}. While compositional mixtures at individual systems are likely to be highly degenerate in most cases, a population study may be able to constrain or rule out a universal background composition.

Uncertainties are often estimated by varying abundances one at a time and inspecting their impact on the quality of the fit. However, it is challenging to quantify all the covariances between the abundances of different species. The Bayesian models used here could have taken advantage of a covariance matrix in the data to further constrain the posterior distributions, but current modelling techniques make this challenging. Efforts should therefore be focused on improving fitting procedures for abundance determination, aiming to produce at least a covariance matrix, but ideally a full description of the posterior distribution such as the output of an MCMC sampler. Ultimately, the aim should be to measure abundances from spectra simultaneously with their compositional analysis, using informative priors, thus minimising degeneracies at every stage, and maximising the scientific return. However, significant challenges must be overcome before that goal can be reached. For example, model grids for fitting multiple elements simultaneously suffer from the curse of dimensionality, as atmosphere models are computationally expensive.

Finally, analyses need not be limited to one star at a time. A hierarchical model can be employed, where characteristics of a population are inferred, and simultaneously used as priors on results for individual stars. Examples of such parameters might include the planetesimal mass function, cooling-age-dependent effects, or the distribution of compositions.

\section{Conclusions}
\label{sectionConclusions}

High-resolution spectra are analysed for seven polluted white dwarfs, measuring metal abundances for up to nine metals. Chondritic compositions are a good match for the material accreted by all seven stars, indicative either of fragments of primitive objects, or that differentiated bodies are being accreted in their entirety, thus averaging out any traces of their internal compositional modifications. One star (J0956) stands out from the sample for being observed well into the decreasing phase, with accretion having ceased several Myr ago. The object that supplied the photospheric metals is inferred to have had a similar mass to the Moon, one of the largest objects yet observed to be polluting a white dwarf.

These results are generated by comparing the data against compositional models based on solar system abundances, where a core--mantle--crust mixture model is found to be competitive with models based on laboratory measurements of meteorites. The accretion histories of each system are inferred or marginalised over, avoiding the need to make assumptions about the state of accretion that would bias the abundance analysis. Recommendations are given for how future work can capitalise on the advantages offered by such Bayesian models.

\section*{Acknowledgements}

The authors are grateful to the anonymous referee, whose comments helped improve the paper. AS thanks Niall Jeffrey and Lorne Whiteway for advice on Bayesian statistics, and Mark Hollands for useful discussions and sharing data. The authors acknowledge support from STFC grants ST/T000406/1 (AS) and ST/R000476/1 (AS, JF, JG), NSFC grant 12203006 (JG) and the Innovation Project of Beijing Academy of Science and Technology (JG; 11000023T000002062763-23CB059). This project has received funding from the European Research Council (ERC) under the European Union’s Horizon 2020 research and innovation programme (Grant agreement No.~101020057). This research made use of \href{https://www.astropy.org}{\textsc{Astropy}} \citep{Astropy2013, Astropy2018}, \href{https://samreay.github.io/ChainConsumer/}{\textsc{ChainConsumer}} \citep{Hinton2016}, \href{https://cmasher.readthedocs.io/}{\textsc{cmasher}} \citep{vanderVelden2020}, \href{https://github.com/joshspeagle/dynesty}{\textsc{dynesty}} \citep{Feroz2009,Higson2019,Speagle2020dynesty}, \textsc{iraf} \citep{Tody1986}, \href{https://www.astro.caltech.edu/~tb/makee/}{\textsc{makee}}, and the \href{https://www.astro.uu.se/valdwiki}{VALD} database, operated at Uppsala University, the Institute of Astronomy RAS in Moscow, and the University of Vienna. The data presented herein were obtained at the W. M. Keck Observatory from telescope time allocated through the National Aeronautics and Space Administration. The Observatory is operated as a scientific partnership among the California Institute of Technology, the University of California and the National Aeronautics and Space Administration. The Observatory was made possible by the generous financial support of the W. M. Keck Foundation.
%

%%%%%%%%%%%%%%%%%%%%%%%%%%%%%%%%%%%%%%%%%%%%%%%%%%
\section*{Data Availability}

The data analysed here are available in the \href{https://koa.ipac.caltech.edu/cgi-bin/KOA/nph-KOAlogin}{Keck Observatory Archive}.



\bibliographystyle{mnras}
\bibliography{DZ_planetesimals_HIRES} % if your bibtex file is called example.bib

\begin{thebibliography}{}
\makeatletter
\relax
\def\mn@urlcharsother{\let\do\@makeother \do\$\do\&\do\#\do\^\do\_\do\%\do\~}
\def\mn@doi{\begingroup\mn@urlcharsother \@ifnextchar [ {\mn@doi@}
  {\mn@doi@[]}}
\def\mn@doi@[#1]#2{\def\@tempa{#1}\ifx\@tempa\@empty \href
  {http://dx.doi.org/#2} {doi:#2}\else \href {http://dx.doi.org/#2} {#1}\fi
  \endgroup}
\def\mn@eprint#1#2{\mn@eprint@#1:#2::\@nil}
\def\mn@eprint@arXiv#1{\href {http://arxiv.org/abs/#1} {{\tt arXiv:#1}}}
\def\mn@eprint@dblp#1{\href {http://dblp.uni-trier.de/rec/bibtex/#1.xml}
  {dblp:#1}}
\def\mn@eprint@#1:#2:#3:#4\@nil{\def\@tempa {#1}\def\@tempb {#2}\def\@tempc
  {#3}\ifx \@tempc \@empty \let \@tempc \@tempb \let \@tempb \@tempa \fi \ifx
  \@tempb \@empty \def\@tempb {arXiv}\fi \@ifundefined
  {mn@eprint@\@tempb}{\@tempb:\@tempc}{\expandafter \expandafter \csname
  mn@eprint@\@tempb\endcsname \expandafter{\@tempc}}}

\bibitem[\protect\citeauthoryear{Adibekyan et~al.,}{Adibekyan
  et~al.}{2021}]{Adibekyan2021}
Adibekyan V.,  et~al., 2021, \mn@doi [Science (80-. ).]
  {10.1126/SCIENCE.ABG8794/SUPPL_FILE/SCIENCE.ABG8794_DATA_S1.ZIP}, 374, 330

\bibitem[\protect\citeauthoryear{Asphaug, Agnor  \& Williams}{Asphaug
  et~al.}{2006}]{Asphaug2006}
Asphaug E.,  Agnor C.~B.,   Williams Q.,  2006, \mn@doi [Nature]
  {10.1038/nature04311}, 439, 155

\bibitem[\protect\citeauthoryear{{Astropy Collaboration} et~al.,}{{Astropy
  Collaboration} et~al.}{2013}]{Astropy2013}
{Astropy Collaboration} et~al., 2013, \mn@doi [A&A]
  {10.1051/0004-6361/201322068}, 558, 33

\bibitem[\protect\citeauthoryear{{Astropy Collaboration} et~al.,}{{Astropy
  Collaboration} et~al.}{2018}]{Astropy2018}
{Astropy Collaboration} et~al., 2018, \mn@doi [Astron. J.]
  {10.3847/1538-3881/aabc4f}, 156, 123

\bibitem[\protect\citeauthoryear{B{\'{e}}dard, Bergeron, Brassard  \&
  Fontaine}{B{\'{e}}dard et~al.}{2020}]{Bedard2020}
B{\'{e}}dard A.,  Bergeron P.,  Brassard P.,   Fontaine G.,  2020, \mn@doi
  [Astrophys. J.] {10.3847/1538-4357/abafbe}, 901, 93

\bibitem[\protect\citeauthoryear{B{\'{e}}dard, Bergeron  \&
  Brassard}{B{\'{e}}dard et~al.}{2023}]{Bedard2023}
B{\'{e}}dard A.,  Bergeron P.,   Brassard P.,  2023, \mn@doi [Astrophys. J.]
  {10.3847/1538-4357/ACBB62}, 946, 24

\bibitem[\protect\citeauthoryear{Bedell et~al.,}{Bedell
  et~al.}{2018}]{Bedell2018}
Bedell M.,  et~al., 2018, \mn@doi [Astrophys. J.] {10.3847/1538-4357/AAD908},
  865, 68

\bibitem[\protect\citeauthoryear{Bergeron, Saumon  \& Wesemael}{Bergeron
  et~al.}{1995}]{Bergeron1995}
Bergeron P.,  Saumon D.,   Wesemael F.,  1995, \mn@doi [Astrophys. J.]
  {10.1086/175566}, 443, 764

\bibitem[\protect\citeauthoryear{Bond, O'Brien  \& Lauretta}{Bond
  et~al.}{2010}]{Bond2010}
Bond J.~C.,  O'Brien D.~P.,   Lauretta D.~S.,  2010, \mn@doi [Astrophys. J.]
  {10.1088/0004-637X/715/2/1050}, 715, 1050

\bibitem[\protect\citeauthoryear{Bonsor, Jofr{\'{e}}, Shorttle, Rogers, Xu  \&
  Melis}{Bonsor et~al.}{2021}]{Bonsor2021}
Bonsor A.,  Jofr{\'{e}} P.,  Shorttle O.,  Rogers L.~K.,  Xu S.,   Melis C.,
  2021, \mn@doi [Mon. Not. R. Astron. Soc.] {10.1093/MNRAS/STAB370}, 503, 1877

\bibitem[\protect\citeauthoryear{Bonsor, Lichtenberg, Drazkowska  \&
  Buchan}{Bonsor et~al.}{2022}]{Bonsor2022}
Bonsor A.,  Lichtenberg T.,  Drazkowska J.,   Buchan A.~M.,  2022, \mn@doi
  [Nat. Astron. 2022 71] {10.1038/s41550-022-01815-8}, 7, 39

\bibitem[\protect\citeauthoryear{Brouwers, Bonsor  \& Malamud}{Brouwers
  et~al.}{2022}]{Brouwers2022asynchronous}
Brouwers M.~G.,  Bonsor A.,   Malamud U.,  2022, \mn@doi [Mon. Not. R. Astron.
  Soc.] {10.1093/MNRAS/STAC3316}, 519, 2646

\bibitem[\protect\citeauthoryear{Buchan, Bonsor, Shorttle, Wade, Harrison,
  Noack  \& Koester}{Buchan et~al.}{2022}]{Buchan2022}
Buchan A.~M.,  Bonsor A.,  Shorttle O.,  Wade J.,  Harrison J.,  Noack L.,
  Koester D.,  2022, \mn@doi [Mon. Not. R. Astron. Soc.]
  {10.1093/MNRAS/STAB3624}, 510, 3512

\bibitem[\protect\citeauthoryear{Chayer, Vennes, Pradhan, Thejll, Beauchamp,
  Fontaine  \& Wesemael}{Chayer et~al.}{1995}]{Chayer1995}
Chayer P.,  Vennes S.,  Pradhan A.~K.,  Thejll P.,  Beauchamp A.,  Fontaine G.,
    Wesemael F.,  1995, \mn@doi [Astrophys. J.] {10.1086/176494}, 454, 429

\bibitem[\protect\citeauthoryear{Collaboration et~al.,}{Collaboration
  et~al.}{2022}]{Gaia2022}
Collaboration G.,  et~al., 2022, \mn@doi [Astron. Astrophys.]
  {10.1051/0004-6361/202243940}, 9, 10

\bibitem[\protect\citeauthoryear{Coutu, Dufour, Bergeron, Blouin, Loranger,
  Allard  \& Dunlap}{Coutu et~al.}{2019}]{Coutu2019}
Coutu S.,  Dufour P.,  Bergeron P.,  Blouin S.,  Loranger E.,  Allard N.~F.,
  Dunlap B.~H.,  2019, \mn@doi [Astrophys. J.] {10.3847/1538-4357/ab46b9}, 885,
  74

\bibitem[\protect\citeauthoryear{Cummings, Kalirai, Tremblay, Ramirez-Ruiz  \&
  Choi}{Cummings et~al.}{2018}]{Cummings2018}
Cummings J.~D.,  Kalirai J.~S.,  Tremblay P.-E.,  Ramirez-Ruiz E.,   Choi J.,
  2018, \mn@doi [Astrophys. J.] {10.3847/1538-4357/aadfd6}, 866, 21

\bibitem[\protect\citeauthoryear{Cunningham et~al.,}{Cunningham
  et~al.}{2021}]{Cunningham2021}
Cunningham T.,  et~al., 2021, \mn@doi [Mon. Not. R. Astron. Soc.]
  {10.1093/mnras/stab553}, 503, 1646

\bibitem[\protect\citeauthoryear{Cunningham, Wheatley, Tremblay,
  G{\"{a}}nsicke, King, Toloza  \& Veras}{Cunningham
  et~al.}{2022}]{Cunningham2022Xrays}
Cunningham T.,  Wheatley P.~J.,  Tremblay P.-E.,  G{\"{a}}nsicke B.~T.,  King
  G.~W.,  Toloza O.,   Veras D.,  2022, \mn@doi [Nat. 2022 6027896]
  {10.1038/s41586-021-04300-w}, 602, 219

\bibitem[\protect\citeauthoryear{Curry, Bonsor, Lichtenberg  \& Shorttle}{Curry
  et~al.}{2022}]{Curry2022}
Curry A.,  Bonsor A.,  Lichtenberg T.,   Shorttle O.,  2022, \mn@doi [MNRAS]
  {10.48550/arxiv.2206.09675}, 000, 1

\bibitem[\protect\citeauthoryear{Dorn, Khan, Heng, Connolly, Alibert, Benz  \&
  Tackley}{Dorn et~al.}{2015}]{Dorn2015}
Dorn C.,  Khan A.,  Heng K.,  Connolly J.~A.,  Alibert Y.,  Benz W.,   Tackley
  P.,  2015, \mn@doi [Astron. Astrophys.] {10.1051/0004-6361/201424915}, 577,
  83

\bibitem[\protect\citeauthoryear{Doyle, Young, Klein, Zuckerman  \&
  Schlichting}{Doyle et~al.}{2019}]{Doyle2019}
Doyle A.~E.,  Young E.~D.,  Klein B.,  Zuckerman B.,   Schlichting H.~E.,
  2019, \mn@doi [Science (80-. ).] {10.1126/science.aax3901}, 366, 356

\bibitem[\protect\citeauthoryear{Doyle, Klein, Schlichting  \& Young}{Doyle
  et~al.}{2020}]{Doyle2020}
Doyle A.~E.,  Klein B.,  Schlichting H.~E.,   Young E.~D.,  2020, \mn@doi
  [Astrophys. J.] {10.3847/1538-4357/abad9a}, 901, 10

\bibitem[\protect\citeauthoryear{Doyle, Desch  \& Young}{Doyle
  et~al.}{2021}]{Doyle2021}
Doyle A.~E.,  Desch S.~J.,   Young E.~D.,  2021, \mn@doi [Astrophys. J.]
  {10.3847/2041-8213/abd9ba}, 907, L35

\bibitem[\protect\citeauthoryear{Dufour et~al.,}{Dufour
  et~al.}{2007}]{Dufour2007}
Dufour P.,  et~al., 2007, \mn@doi [Astrophys. J.] {10.1086/518468}, 663, 1291

\bibitem[\protect\citeauthoryear{Dufour, Kidlic, Fontaine, Bergeron, Melis  \&
  Bochanski}{Dufour et~al.}{2012}]{Dufour2012}
Dufour P.,  Kidlic M.,  Fontaine G.,  Bergeron P.,  Melis C.,   Bochanski J.,
  2012, \mn@doi [Astrophys. J.] {10.1088/0004-637X/749/1/6}, 749

\bibitem[\protect\citeauthoryear{Dumusque}{Dumusque}{2018}]{Dumusque2018}
Dumusque X.,  2018, \mn@doi [Astron. Astrophys.] {10.1051/0004-6361/201833795},
  620, 47

\bibitem[\protect\citeauthoryear{Eisenstein et~al.,}{Eisenstein
  et~al.}{2006}]{Eisenstein2006}
Eisenstein D.~J.,  et~al., 2006, \mn@doi [Astrophys. J. Suppl. Ser.]
  {10.1086/507110}, 167, 40

\bibitem[\protect\citeauthoryear{Elms et~al.,}{Elms et~al.}{2022}]{Elms2022}
Elms A.~K.,  et~al., 2022, \mn@doi [MNRAS] {10.48550/arxiv.2206.05258}, 000, 1

\bibitem[\protect\citeauthoryear{Farihi}{Farihi}{2016}]{Farihi2016}
Farihi J.,  2016, \mn@doi [New Astron. Rev.] {10.1016/j.newar.2016.03.001}, 71,
  9

\bibitem[\protect\citeauthoryear{Farihi, Barstow, Redfield, Dufour  \&
  Hambly}{Farihi et~al.}{2010}]{Farihi2010rockyPlanetesimals}
Farihi J.,  Barstow M.~A.,  Redfield S.,  Dufour P.,   Hambly N.~C.,  2010,
  \mn@doi [Mon. Not. R. Astron. Soc.] {10.1111/j.1365-2966.2010.16426.x}, 404,
  2123

\bibitem[\protect\citeauthoryear{Farihi, G{\"{a}}nsicke, Wyatt, Girven, Pringle
   \& King}{Farihi et~al.}{2012}]{Farihi2012intenseAccretion}
Farihi J.,  G{\"{a}}nsicke B.~T.,  Wyatt M.~C.,  Girven J.,  Pringle J.~E.,
  King A.~R.,  2012, \mn@doi [Mon. Not. R. Astron. Soc.]
  {10.1111/j.1365-2966.2012.21215.x}, 424, 464

\bibitem[\protect\citeauthoryear{Farihi, G{\"{a}}nsicke  \& Koester}{Farihi
  et~al.}{2013a}]{Farihi2013}
Farihi J.,  G{\"{a}}nsicke B.~T.,   Koester D.,  2013a, \mn@doi [Science (80-.
  ).] {10.1126/science.1239447}, 342, 218

\bibitem[\protect\citeauthoryear{Farihi, Bond, Dufour, Haghighipour, Schaefer,
  Holberg, Barstow  \& Burleigh}{Farihi et~al.}{2013b}]{Farihi2013Hubble}
Farihi J.,  Bond H.~E.,  Dufour P.,  Haghighipour N.,  Schaefer G.~H.,  Holberg
  J.~B.,  Barstow M.~A.,   Burleigh M.~R.,  2013b, \mn@doi [Mon. Not. R.
  Astron. Soc.] {10.1093/MNRAS/STS677}, 430, 652

\bibitem[\protect\citeauthoryear{Farihi et~al.,}{Farihi
  et~al.}{2018}]{Farihi2018Magnetism}
Farihi J.,  et~al., 2018, \mn@doi [Mon. Not. R. Astron. Soc.]
  {10.1093/mnras/stx2664}, 474, 947

\bibitem[\protect\citeauthoryear{Farihi et~al.,}{Farihi
  et~al.}{2022}]{Farihi2022}
Farihi J.,  et~al., 2022, \mn@doi [Mon. Not. R. Astron. Soc.]
  {10.1093/MNRAS/STAB3475}, 511, 1647

\bibitem[\protect\citeauthoryear{Feroz, Hobson  \& Bridges}{Feroz
  et~al.}{2009}]{Feroz2009}
Feroz F.,  Hobson M.~P.,   Bridges M.,  2009, \mn@doi [Mon. Not. R. Astron.
  Soc.] {10.1111/J.1365-2966.2009.14548.X/2/M_MNRAS0398-1601-M32.GIF}, 398,
  1601

\bibitem[\protect\citeauthoryear{G{\"{a}}nsicke, Koester, Farihi, Girven,
  Parsons  \& Breedt}{G{\"{a}}nsicke et~al.}{2012}]{Gansicke2012}
G{\"{a}}nsicke B.~T.,  Koester D.,  Farihi J.,  Girven J.,  Parsons S.~G.,
  Breedt E.,  2012, \mn@doi [Mon. Not. R. Astron. Soc.]
  {10.1111/j.1365-2966.2012.21201.x}, 424, 333

\bibitem[\protect\citeauthoryear{G{\"{a}}nsicke, Schreiber, Toloza, Fusillo,
  Koester  \& Manser}{G{\"{a}}nsicke et~al.}{2019}]{Gansicke2019}
G{\"{a}}nsicke B.~T.,  Schreiber M.~R.,  Toloza O.,  Fusillo N. P.~G.,  Koester
  D.,   Manser C.~J.,  2019, \mn@doi [Nature] {10.1038/s41586-019-1789-8}, 576,
  61

\bibitem[\protect\citeauthoryear{Gillon et~al.,}{Gillon
  et~al.}{2017}]{Gillon2017}
Gillon M.,  et~al., 2017, \mn@doi [Nature] {10.1038/nature21360}, 542, 456

\bibitem[\protect\citeauthoryear{Girven, Brinkworth, Farihi, G{\"{a}}nsicke,
  Hoard, Marsh  \& Koester}{Girven et~al.}{2012}]{Girven2012}
Girven J.,  Brinkworth C.~S.,  Farihi J.,  G{\"{a}}nsicke B.~T.,  Hoard D.~W.,
  Marsh T.~R.,   Koester D.,  2012, \mn@doi [Astrophys. J.]
  {10.1088/0004-637X/749/2/154}, 749, 154

\bibitem[\protect\citeauthoryear{Guidry et~al.,}{Guidry
  et~al.}{2021}]{Guidry2021}
Guidry J.~A.,  et~al., 2021, \mn@doi [Astrophys. J.]
  {10.3847/1538-4357/abee68}, 912, 125

\bibitem[\protect\citeauthoryear{Harrison et~al.,}{Harrison
  et~al.}{2021a}]{Harrison2021bayesian}
Harrison J. H.~D.,  et~al., 2021a, \mn@doi [Mon. Not. R. Astron. Soc.]
  {10.1093/mnras/stab736}, 504, 2853

\bibitem[\protect\citeauthoryear{Harrison, Shorttle  \& Bonsor}{Harrison
  et~al.}{2021b}]{Harrison2021postNebula}
Harrison J.~H.,  Shorttle O.,   Bonsor A.,  2021b, \mn@doi [Earth Planet. Sci.
  Lett.] {10.1016/j.epsl.2020.116694}, 554, 116694

\bibitem[\protect\citeauthoryear{Higson, Handley, Hobson  \& Lasenby}{Higson
  et~al.}{2019}]{Higson2019}
Higson E.,  Handley W.,  Hobson M.,   Lasenby A.,  2019, \mn@doi [Stat.
  Comput.] {10.1007/S11222-018-9844-0/TABLES/12}, 29, 891

\bibitem[\protect\citeauthoryear{Hinton}{Hinton}{2016}]{Hinton2016}
Hinton S.,  2016, \mn@doi [J. Open Source Softw.] {10.21105/JOSS.00045}, 1, 45

\bibitem[\protect\citeauthoryear{Hollands, Koester, Alekseev, Herbert  \&
  G{\"{a}}nsicke}{Hollands et~al.}{2017}]{Hollands2017abundances}
Hollands M.~A.,  Koester D.,  Alekseev V.,  Herbert E.~L.,   G{\"{a}}nsicke
  B.~T.,  2017, \mn@doi [Mon. Not. R. Astron. Soc.] {10.1093/mnras/stx250},
  467, stx250

\bibitem[\protect\citeauthoryear{Hollands, G{\"{a}}nsicke  \& Koester}{Hollands
  et~al.}{2018}]{Hollands2018analysis}
Hollands M.~A.,  G{\"{a}}nsicke B.~T.,   Koester D.,  2018, \mn@doi [Mon. Not.
  R. Astron. Soc.] {10.1093/mnras/sty592}, 477, 93

\bibitem[\protect\citeauthoryear{Hollands, Tremblay, G{\"{a}}nsicke, Koester
  \& Gentile-Fusillo}{Hollands et~al.}{2021}]{Hollands2021alkali}
Hollands M.~A.,  Tremblay P.~E.,  G{\"{a}}nsicke B.~T.,  Koester D.,
  Gentile-Fusillo N.~P.,  2021, \mn@doi [Nat. Astron.]
  {10.1038/s41550-020-01296-7}, 5, 451

\bibitem[\protect\citeauthoryear{Hollands, Tremblay, G{\"{a}}nsicke  \&
  Koester}{Hollands et~al.}{2022}]{Hollands2022coolWhiteDwarfs}
Hollands M.~A.,  Tremblay P.~E.,  G{\"{a}}nsicke B.~T.,   Koester D.,  2022,
  \mn@doi [Mon. Not. R. Astron. Soc.] {10.1093/MNRAS/STAB3696}, 511, 71

\bibitem[\protect\citeauthoryear{Hoskin et~al.,}{Hoskin
  et~al.}{2020}]{Hoskin2020}
Hoskin M.~J.,  et~al., 2020, \mn@doi [Mon. Not. R. Astron. Soc.]
  {10.1093/mnras/staa2717}, 499, 171

\bibitem[\protect\citeauthoryear{Izquierdo, Toloza, G{\"{a}}nsicke,
  Rodr{\'{i}}guez-Gil, Farihi, Koester, Guo  \& Redfield}{Izquierdo
  et~al.}{2021}]{Izquierdo2021GD424}
Izquierdo P.,  Toloza O.,  G{\"{a}}nsicke B.~T.,  Rodr{\'{i}}guez-Gil P.,
  Farihi J.,  Koester D.,  Guo J.,   Redfield S.,  2021, \mn@doi [Mon. Not. R.
  Astron. Soc.] {10.1093/MNRAS/STAA3987}, 501, 4276

\bibitem[\protect\citeauthoryear{Izquierdo, G, Rodr{\'{i}}guez-Gil, Koester,
  Toloza, {Gentile Fusillo}, Pala  \& Tremblay}{Izquierdo
  et~al.}{2023}]{Izquierdo2023}
Izquierdo P.,  G B.~T.,  Rodr{\'{i}}guez-Gil P.,  Koester D.,  Toloza O.,
  {Gentile Fusillo} N.~P.,  Pala A.~F.,   Tremblay P.-E.,  2023, \mn@doi [Mon.
  Not. R. Astron. Soc.] {10.1093/MNRAS/STAD282}, 520, 2843

\bibitem[\protect\citeauthoryear{Johnson, Klein, Koester, Melis, Zuckerman  \&
  Jura}{Johnson et~al.}{2022}]{Johnson2023}
Johnson T.~M.,  Klein B.~L.,  Koester D.,  Melis C.,  Zuckerman B.,   Jura M.,
  2022, \mn@doi [Astrophys. J.] {10.3847/1538-4357/ACA089}, 941, 113

\bibitem[\protect\citeauthoryear{Jura, Farihi  \& Zuckerman}{Jura
  et~al.}{2007}]{Jura2007Spitzer11stars}
Jura M.,  Farihi J.,   Zuckerman B.,  2007, \mn@doi [Astrophys. J.]
  {10.1086/518767}, 663, 1285

\bibitem[\protect\citeauthoryear{Jura, Farihi  \& Zuckerman}{Jura
  et~al.}{2009}]{Jura2009silicates}
Jura M.,  Farihi J.,   Zuckerman B.,  2009, \mn@doi [Astron. J.]
  {10.1088/0004-6256/137/2/3191}, 137, 3191

\bibitem[\protect\citeauthoryear{Jura, Xu  \& Young}{Jura
  et~al.}{2013}]{Jura2013}
Jura M.,  Xu S.,   Young E.~D.,  2013, \mn@doi [Astrophys. J. Lett.]
  {10.1088/2041-8205/775/2/L41}, 775, L41

\bibitem[\protect\citeauthoryear{Klein \& Moeschberger}{Klein \&
  Moeschberger}{2003}]{Klein2003survival}
Klein J.~P.,  Moeschberger M.~L.,  2003, {Survival Analysis}.
Statistics for Biology and Health, Springer New York, New York, NY,
  \mn@doi{10.1007/b97377}

\bibitem[\protect\citeauthoryear{Klein, Jura, Koester, Zuckerman  \&
  Melis}{Klein et~al.}{2010}]{Klein2010}
Klein B.,  Jura M.,  Koester D.,  Zuckerman B.,   Melis C.,  2010, \mn@doi
  [Astrophys. J.] {10.1088/0004-637X/709/2/950}, 709, 950

\bibitem[\protect\citeauthoryear{Klein, Doyle, Zuckerman, Dufour, Blouin,
  Melis, Weinberger  \& Young}{Klein et~al.}{2021}]{Klein2021}
Klein B.~L.,  Doyle A.~E.,  Zuckerman B.,  Dufour P.,  Blouin S.,  Melis C.,
  Weinberger A.~J.,   Young E.~D.,  2021, \mn@doi [Astrophys. J.]
  {10.3847/1538-4357/ABE40B}, 914, 61

\bibitem[\protect\citeauthoryear{Koester}{Koester}{2009}]{Koester2009}
Koester D.,  2009, \mn@doi [Astron. Astrophys.] {10.1051/0004-6361/200811468},
  498, 517

\bibitem[\protect\citeauthoryear{Koester, G{\"{a}}nsicke  \& Farihi}{Koester
  et~al.}{2014}]{Koester2014}
Koester D.,  G{\"{a}}nsicke B.~T.,   Farihi J.,  2014, \mn@doi [Astron.
  Astrophys.] {10.1051/0004-6361/201423691}, 566, A34

\bibitem[\protect\citeauthoryear{Koester, Kepler  \& Irwin}{Koester
  et~al.}{2020}]{Koester2020atmospheres}
Koester D.,  Kepler S.~O.,   Irwin A.~W.,  2020, \mn@doi [Astron. Astrophys.]
  {10.1051/0004-6361/202037530}, 635, 103

\bibitem[\protect\citeauthoryear{Lisse et~al.,}{Lisse et~al.}{2019}]{Lisse2019}
Lisse C.~M.,  et~al., 2019, \mn@doi [Res. Notes AAS]
  {10.3847/2515-5172/AB2E0E}, 3, 90

\bibitem[\protect\citeauthoryear{Lodders}{Lodders}{2003}]{Lodders2003}
Lodders K.,  2003, \mn@doi [Astrophys. J.] {10.1086/375492}, 591, 1220

\bibitem[\protect\citeauthoryear{Malamud \& Perets}{Malamud \&
  Perets}{2020}]{Malamud2020A}
Malamud U.,  Perets H.~B.,  2020, \mn@doi [Mon. Not. R. Astron. Soc.]
  {10.1093/mnras/staa142}, 492, 5561

\bibitem[\protect\citeauthoryear{Monnereau, Guignard, N{\'{e}}ri, Toplis  \&
  Quitt{\'{e}}}{Monnereau et~al.}{2023}]{Monnereau2023}
Monnereau M.,  Guignard J.,  N{\'{e}}ri A.,  Toplis M.~J.,   Quitt{\'{e}} G.,
  2023, \mn@doi [Icarus] {10.1016/J.ICARUS.2022.115294}, 390, 115294

\bibitem[\protect\citeauthoryear{Moskovitz \& Gaidos}{Moskovitz \&
  Gaidos}{2011}]{Moskovitz2011}
Moskovitz N.,  Gaidos E.,  2011, \mn@doi [Meteorit. Planet. Sci.]
  {10.1111/j.1945-5100.2011.01201.x}, 46, 903

\bibitem[\protect\citeauthoryear{Nittler, McCoy, Clark, Murphy, Trombka  \&
  Jarosewich}{Nittler et~al.}{2004}]{Nittler2004}
Nittler L.~R.,  McCoy T.~J.,  Clark P.~E.,  Murphy M.~E.,  Trombka J.~I.,
  Jarosewich E.,  2004, Antarct. Meteor. Res., 17, 231

\bibitem[\protect\citeauthoryear{Nittler, Chabot, Grove  \& Peplowski}{Nittler
  et~al.}{2018}]{Nittler2018}
Nittler L.~R.,  Chabot N.~L.,  Grove T.~L.,   Peplowski P.~N.,  2018, in
  Solomon S.~C.,  Nittler L.~R.,   Anderson B.~J.,  eds, , Mercur. View after
  Messenger.
Cambridge University Press, Chapt.~2, pp 30--51,
  \mn@doi{10.1017/9781316650684.003}

\bibitem[\protect\citeauthoryear{Piskunov, Kupka, Ryabchikova, Weiss  \&
  Jeffery}{Piskunov et~al.}{1995}]{Piskunov1995}
Piskunov N.~E.,  Kupka F.,  Ryabchikova T.~A.,  Weiss W.~W.,   Jeffery C.~S.,
  1995, Astron. Astrophys. Suppl. Ser., 112, 525

\bibitem[\protect\citeauthoryear{Putirka \& Xu}{Putirka \&
  Xu}{2021}]{Putirka2021}
Putirka K.~D.,  Xu S.,  2021, \mn@doi [Nat. Commun. 2021 121]
  {10.1038/s41467-021-26403-8}, 12, 1

\bibitem[\protect\citeauthoryear{Raddi, G{\"{a}}nsicke, Koester, Farihi,
  Hermes, Scaringi, Breedt  \& Girven}{Raddi et~al.}{2015}]{Raddi2015}
Raddi R.,  G{\"{a}}nsicke B.~T.,  Koester D.,  Farihi J.,  Hermes J.~J.,
  Scaringi S.,  Breedt E.,   Girven J.,  2015, \mn@doi [Mon. Not. R. Astron.
  Soc.] {10.1093/mnras/stv701}, 450, 2083

\bibitem[\protect\citeauthoryear{Rafikov}{Rafikov}{2011}]{Rafikov2011runaway}
Rafikov R.~R.,  2011, \mn@doi [Mon. Not. R. Astron. Soc.]
  {10.1111/j.1745-3933.2011.01096.x}, 416, L55

\bibitem[\protect\citeauthoryear{Rolland, Bergeron  \& Fontaine}{Rolland
  et~al.}{2018}]{Rolland2018}
Rolland B.,  Bergeron P.,   Fontaine G.,  2018, \mn@doi [Astrophys. J.]
  {10.3847/1538-4357/aab713}, 857, 56

\bibitem[\protect\citeauthoryear{Rubin et~al.,}{Rubin et~al.}{2019}]{Rubin2019}
Rubin M.,  et~al., 2019, \mn@doi [Mon. Not. R. Astron. Soc.]
  {10.1093/mnras/stz2086}, 489, 594

\bibitem[\protect\citeauthoryear{Rumble}{Rumble}{2019}]{Rumble2019CRChandbook}
Rumble J.~R.,  ed. 2019, {CRC handbook of chemistry and physics}, 100 edn.
CRC Press, Boca Raton, Fla.

\bibitem[\protect\citeauthoryear{Schatzman}{Schatzman}{1945}]{Schatzman1945}
Schatzman E.,  1945, Ann. d'Astrophysique, 8, 143

\bibitem[\protect\citeauthoryear{Shah, Helled, Alibert  \& Mezger}{Shah
  et~al.}{2022}]{Shah2022}
Shah O.,  Helled R.,  Alibert Y.,   Mezger K.,  2022, \mn@doi [Astrophys. J.]
  {10.3847/1538-4357/AC410D}, 926, 217

\bibitem[\protect\citeauthoryear{Skilling}{Skilling}{2004}]{Skilling2004}
Skilling J.,  2004, in J. Chem. Phys.. AIP Publishing, pp 395--405,
  \mn@doi{10.1063/1.1835238}

\bibitem[\protect\citeauthoryear{Skilling}{Skilling}{2006}]{Skilling2006}
Skilling J.,  2006, \mn@doi [https://doi.org/10.1214/06-BA127]
  {10.1214/06-BA127}, 1, 833

\bibitem[\protect\citeauthoryear{Speagle}{Speagle}{2020}]{Speagle2020dynesty}
Speagle J.~S.,  2020, \mn@doi [Mon. Not. R. Astron. Soc.]
  {10.1093/mnras/staa278}, 493, 3132

\bibitem[\protect\citeauthoryear{Swan, Farihi, Koester, Hollands, Parsons,
  Cauley, Redfield  \& G{\"{a}}nsicke}{Swan et~al.}{2019}]{Swan2019Xshooter}
Swan A.,  Farihi J.,  Koester D.,  Hollands M.,  Parsons S.,  Cauley P.~W.,
  Redfield S.,   G{\"{a}}nsicke B.~T.,  2019, \mn@doi [Mon. Not. R. Astron.
  Soc.] {10.1093/mnras/stz2337}, 490, 202

\bibitem[\protect\citeauthoryear{Taylor \& McLennan}{Taylor \&
  McLennan}{2008}]{Taylor2008PlanetaryCrusts}
Taylor S.~R.,  McLennan S.,  2008, {Planetary Crusts}.
Cambridge University Press, Cambridge

\bibitem[\protect\citeauthoryear{Tody}{Tody}{1986}]{Tody1986}
Tody D.,  1986, in Instrum. Astron. VI. SPIE, p.~733,
  \mn@doi{10.1117/12.968154}

\bibitem[\protect\citeauthoryear{Trierweiler, Doyle  \& Young}{Trierweiler
  et~al.}{2023}]{Trierweiler2023}
Trierweiler I.~L.,  Doyle A.~E.,   Young E.~D.,  2023, \mn@doi [Planet. Sci.
  J.] {10.3847/PSJ/ACDEF3}, 4, 136

\bibitem[\protect\citeauthoryear{Trotta}{Trotta}{2008}]{Trotta2008}
Trotta R.,  2008, \mn@doi [Contemp. Phys.] {10.1080/00107510802066753}, 49, 71

\bibitem[\protect\citeauthoryear{Turner \& Wyatt}{Turner \&
  Wyatt}{2020}]{Turner2020}
Turner S. G.~D.,  Wyatt M.~C.,  2020, \mn@doi [MNRAS] {10.1093/mnras/stz3191},
  491, 4672

\bibitem[\protect\citeauthoryear{Unterborn, Desch, Hinkel  \&
  Lorenzo}{Unterborn et~al.}{2018}]{Unterborn2018}
Unterborn C.~T.,  Desch S.~J.,  Hinkel N.~R.,   Lorenzo A.,  2018, \mn@doi
  [Nat. Astron. 2018 24] {10.1038/s41550-018-0411-6}, 2, 297

\bibitem[\protect\citeauthoryear{Vanderburg et~al.,}{Vanderburg
  et~al.}{2015}]{vanderburg2015}
Vanderburg A.,  et~al., 2015, \mn@doi [Nature] {10.1038/nature15527}, 526, 546

\bibitem[\protect\citeauthoryear{{\noopsort{Vandervelden}}van~der
  Velden}{{\noopsort{Vandervelden}}van~der Velden}{2020}]{vanderVelden2020}
{\noopsort{Vandervelden}}van~der Velden E.,  2020, \mn@doi [J. Open Source
  Softw.] {10.21105/joss.02004}, 5, 2004

\bibitem[\protect\citeauthoryear{{\noopsort{Vanmaanen}}van~Maanen}{{\noopsort{Vanmaanen}}van~Maanen}{1917}]{vanMaanen1917}
{\noopsort{Vanmaanen}}van~Maanen A.,  1917, \mn@doi [Publ. Astron. Soc.
  Pacific] {10.1086/122654}, 29, 258

\bibitem[\protect\citeauthoryear{Vogt et~al.,}{Vogt et~al.}{1994}]{Vogt1994}
Vogt S.~S.,  et~al., 1994, in Instrum. Astron. VIII. SPIE, pp 362--375,
  \mn@doi{10.1117/12.176725}

\bibitem[\protect\citeauthoryear{Wang, Lineweaver  \& Ireland}{Wang
  et~al.}{2018}]{Wang2018earth}
Wang H.~S.,  Lineweaver C.~H.,   Ireland T.~R.,  2018, \mn@doi [Icarus]
  {10.1016/j.icarus.2017.08.024}, 299, 460

\bibitem[\protect\citeauthoryear{Wang, Lineweaver  \& Ireland}{Wang
  et~al.}{2019}]{Wang2019protosolar}
Wang H.~S.,  Lineweaver C.~H.,   Ireland T.~R.,  2019, \mn@doi [Icarus]
  {10.1016/J.ICARUS.2019.03.018}, 328, 287

\bibitem[\protect\citeauthoryear{Weisberg, McCoy  \& Krot}{Weisberg
  et~al.}{2006}]{Weisberg2006}
Weisberg M.~K.,  McCoy T.~J.,   Krot A.~N.,  2006, in Lauretta D.~S.,  McSween
  H.~Y.,  eds, , Meteorites Early Sol. Syst. II.
University of Arizona Press, Tuscon, p.~19

\bibitem[\protect\citeauthoryear{Wilson, Farihi, G{\"{a}}nsicke  \&
  Swan}{Wilson et~al.}{2019}]{Wilson2019}
Wilson T.~G.,  Farihi J.,  G{\"{a}}nsicke B.~T.,   Swan A.,  2019, \mn@doi
  [Mon. Not. R. Astron. Soc.] {10.1093/mnras/stz1050}, 487, 133

\bibitem[\protect\citeauthoryear{Wolszczan \& Frail}{Wolszczan \&
  Frail}{1992}]{Wolszczan1992}
Wolszczan A.,  Frail D.~A.,  1992, \mn@doi [Nature] {10.1038/355145a0}, 355,
  145

\bibitem[\protect\citeauthoryear{Wyatt, Farihi, Pringle  \& Bonsor}{Wyatt
  et~al.}{2014}]{Wyatt2014}
Wyatt M.~C.,  Farihi J.,  Pringle J.~E.,   Bonsor A.,  2014, \mn@doi [Mon. Not.
  R. Astron. Soc.] {10.1093/mnras/stu183}, 439, 3371

\bibitem[\protect\citeauthoryear{Xu, Jura, Klein, Koester  \& Zuckerman}{Xu
  et~al.}{2013}]{Xu2013}
Xu S.,  Jura M.,  Klein B.,  Koester D.,   Zuckerman B.,  2013, \mn@doi
  [Astrophys. J.] {10.1088/0004-637X/766/2/132}, 766, 132

\bibitem[\protect\citeauthoryear{Xu, Zuckerman, Dufour, Young, Klein  \&
  Jura}{Xu et~al.}{2017}]{Xu2017}
Xu S.,  Zuckerman B.,  Dufour P.,  Young E.~D.,  Klein B.,   Jura M.,  2017,
  \mn@doi [Astrophys. J.] {10.3847/2041-8213/836/1/L7}, 836, L7

\bibitem[\protect\citeauthoryear{Xu, Dufour, Klein, Melis, Monson, Zuckerman,
  Young  \& Jura}{Xu et~al.}{2019}]{Xu2019composition}
Xu S.,  Dufour P.,  Klein B.,  Melis C.,  Monson N.~N.,  Zuckerman B.,  Young
  E.~D.,   Jura M.~A.,  2019, \mn@doi [Astron. J.] {10.3847/1538-3881/ab4cee},
  158, 242

\bibitem[\protect\citeauthoryear{York et~al.,}{York
  et~al.}{2000}]{York2000SDSS}
York D.~G.,  et~al., 2000, \mn@doi [Astron. J.] {10.1086/301513}, 120, 1579

\bibitem[\protect\citeauthoryear{Yoshizaki \& McDonough}{Yoshizaki \&
  McDonough}{2020}]{Yoshizaki2020Mars}
Yoshizaki T.,  McDonough W.~F.,  2020, \mn@doi [Geochim. Cosmochim. Acta]
  {10.1016/j.gca.2020.01.011}, 273, 137

\bibitem[\protect\citeauthoryear{Yoshizaki \& McDonough}{Yoshizaki \&
  McDonough}{2021}]{Yoshizaki2021Mars}
Yoshizaki T.,  McDonough W.~F.,  2021, \mn@doi [Geochemistry]
  {10.1016/J.CHEMER.2021.125746}, 81, 125746

\bibitem[\protect\citeauthoryear{Zuckerman \& Becklin}{Zuckerman \&
  Becklin}{1987}]{Zuckerman1987}
Zuckerman B.,  Becklin E.~E.,  1987, \mn@doi [Nature] {10.1038/330138a0}, 330,
  138

\bibitem[\protect\citeauthoryear{Zuckerman, Koester, Melis, Hansen  \&
  Jura}{Zuckerman et~al.}{2007}]{Zuckerman2007}
Zuckerman B.,  Koester D.,  Melis C.,  Hansen B.~M.,   Jura M.,  2007, \mn@doi
  [Astrophys. J.] {10.1086/522223}, 671, 872

\bibitem[\protect\citeauthoryear{Zuckerman, Melis, Klein, Koester  \&
  Jura}{Zuckerman et~al.}{2010}]{Zuckerman2010}
Zuckerman B.,  Melis C.,  Klein B.,  Koester D.,   Jura M.,  2010, \mn@doi
  [Astrophys. J.] {10.1088/0004-637X/722/1/725}, 722, 725

\bibitem[\protect\citeauthoryear{Zuckerman, Koester, Dufour, Melis, Klein  \&
  Jura}{Zuckerman et~al.}{2011}]{Zuckerman2011}
Zuckerman B.,  Koester D.,  Dufour P.,  Melis C.,  Klein B.,   Jura M.,  2011,
  \mn@doi [Astrophys. J.] {10.1088/0004-637X/739/2/101}, 739, 101

\makeatother
\end{thebibliography}

%\begin{thebibliography}{99}
%\end{thebibliography}

%%%%%%%%%%%%%%%%%%%%%%%%%%%%%%%%%%%%%%%%%%%%%%%%%%

%%%%%%%%%%%%%%%%% APPENDICES %%%%%%%%%%%%%%%%%%%%%
\clearpage

\appendix
\section{Bayesian framework for accretion events}
\label{appendixBayesian}

\subsection{Model specification}
\label{subsectionModelSpecification}

\begin{table}
\begin{center}
\caption{Glossary of symbols. Boldface indicates a vector. Generic terms feature in the discussion on Bayesian analysis.}
\label{tableSymbols}
%\resizebox{\textwidth}{!}{ %If it's slightly too wide
\begin{tabular}{ll}
\hline
Symbol & Meaning \\
\hline

$B_{\text{ab}}$        & Bayes factor of models $\mathcal{M}_{\text{a}}$ and $\mathcal{M}_{\text{b}}$ \\
$f_x$                  & Fraction of component $x$ in mixture model \\
$g$                    & Stellar surface gravity \\
$\mathcal{L}$          & Likelihood function (generic) \\
$M_{\text{acc}}$       & Mass of accreted material \\
$\dot{M}_{\text{acc}}$ & Accretion rate \\
$M_{\textrm{cvz}}$     & Convection zone mass \\
$\mathcal{M}$          & Model (generic) \\
$O_{\text{ab}}$        & Posterior odds of models $\mathcal{M}_{\text{a}}$ and $\mathcal{M}_{\text{b}}$ \\
$p(\mathcal{M})$       & Prior probabilty on a model (generic) \\
$P(\bm{\theta})$       & Posterior probability distributions of parameters $\bm{\theta}$ (generic) \\
$S$                    & Survival function (generic) \\
$t_1$                  & Duration of accretion episode \\
$t_2$                  & Time since accretion ceased \\
$\teff$                & Stellar effective temperature \\
$\bm{x}$               & Data (generic) \\
$\bm{X}$               & Observed photospheric abundances (numerical, relative to helium) \\
$\bm{X}_\textrm{a}$    & Elemental mass fractions in accreted material \\
$X_{\textrm{H},0}$     & Photospheric hydrogen abundance before accretion episode \\
$\bm{X}_\textrm{m}$    & Photospheric abundances predicted by model \\
$\mathcal{Z}$          & Bayesian evidence \\
$\delta$               & Switching parameter \\
$\bm{\theta}$          & Parameters of a model (generic)\\
$\pi(\bm{\theta})$     & Priors on parameters $\bm{\theta}$ (generic) \\
$\bm{\sigma}$          & Uncertainties on data (generic) \\
$\sigma_k$             & Uncertainty on observed photospheric abundance $X_k$ of element $k$ \\
$\tau_k$               & Sinking timescale of element $k$ \\

\hline
\end{tabular}
%} %Closing brace for \resizebox, if we need it
\end{center}
\end{table}

The general equation for the photospheric mass abundance $X_{\text{m},k}$ of a metal $k$ that is accreted at a rate $\dot{M}_{\text{acc},k}$ for a time $t$ is (\citealt{Koester2009}, equation~5):

\begin{equation}
\label{equationAccretionGeneral}
    X_{\text{m},k}=X_{\text{m},k,0}\,e^{-t/\tau_k}+\frac{\tau_k\dot{M}_{\text{acc},k}}{M_{\text{cvz}}}(1-e^{-t/\tau_k})
\end{equation}

The accretion rate for a given metal is $\dot{M}_{\text{acc},k}=X_{\text{a},k}\dot{M}_{\text{acc}}$, where $\dot{M}_{\text{acc}}$ is the total accretion rate, and $X_{\text{a},k}$ is the mass fraction of element $k$ in the accreted material. White dwarf pollution data are normally presented as numerical abundances of photospheric metals relative to the dominant element (hydrogen or helium), including those in Section~\ref{sectionObservationsAndModelling}. However, the model is specified in terms of mass fractions in the accreted material, so care must be taken not to confuse the two representations.

Where the atmosphere initially consists of only hydrogen or helium, the initial photospheric abundances of all metals $k$ are $X_{\text{m},k,0}=0$. From this starting point, for accretion that proceeds for a time $t_1$, Equation~\ref{equationAccretionGeneral} reduces to:

\begin{equation}
\label{equationAccretionT1}
    X_{\text{m},k}=\frac{\tau_k\dot{M}_{\text{acc},k}}{M_{\text{cvz}}}(1-e^{-t_1/\tau_k})
\end{equation}

If accretion then ceases, and a time $t_2$ passes before the star is observed, Equation~\ref{equationAccretionGeneral} is evaluated with $\dot{M}_{\text{acc},k}=0$, using Equation~\ref{equationAccretionT1} to set $X_{\text{m},k,0}$:

\begin{equation}
\label{equationAccretionT2}
    X_{\text{m},k}=X_{\text{m},k,0}\,e^{-t_2/\tau_k}
\end{equation}

For simplicity, $\tau$ is assumed to remain constant, despite being a function of $X_m$, as the largest variation of $X_m$ occurs during the brief increasing phase.

\begin{figure}
\includegraphics[width=\columnwidth]{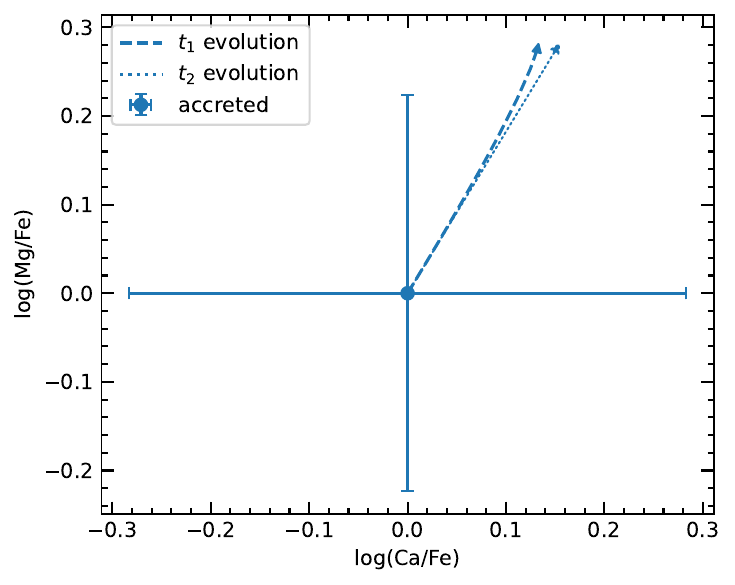}
\caption{Simulated evolution of photospheric abundance ratios during and after an accretion event that proceeds at a constant rate. The dashed line represents accretion moving from the increasing phase to a steady state. The dotted line shows the effect of moving $1.0\tau_{\text{Ca}}$ into the decreasing phase, starting from the increasing phase. The two effects are highly degenerate with each other in this two-dimensional case, and the difference is comparable to observational uncertainties. Sinking timescales and error bars from J0956 are used to generate a realistic example, but accreted abundance ratios are set to unity to avoid confusion with real data.}
\label{figureT1T2degeneracy}
\end{figure}

Equations~\ref{equationAccretionT1} and~\ref{equationAccretionT2} drive abundance ratios in a similar -- but not identical -- direction during and after accretion, as illustrated in Fig.~\ref{figureT1T2degeneracy}. There is therefore some degeneracy between $t_1$ and $t_2$, as the change in abundance ratios between the increasing phase and steady state is similar to that between $t_2=0$ and $t_2=\tau_{\text{Ca}}$ in the decreasing phase. While it may appear futile to attempt to distinguish these effects on a two-dimensional plot, differences widen in a higher-dimensional space when more metal species are available.

As outlined above, the main parameters of an accretion episode are $\dot{M}_{\text{acc}}$, $t_1$, and $t_2$. As $M_{\text{acc}}=\dot{M}_{\text{acc}}t_1$, it is possible to use $M_{\text{acc}}$ as a parameter instead of $\dot{M}_{\text{acc}}$, subject to a degeneracy with $t_1$ when accretion approaches a steady state. For hydrogen-dominated stars, $\dot{M}_{\text{acc}}$ would be a more appropriate choice, but the long sinking timescales of the DZ~stars studied here make it reasonable to attempt to infer $M_{\text{acc}}$ directly. Stellar $\teff$ and $\log{g}$ are secondary parameters, as they determine $M_{\text{cvz}}$ and $\tau$, and therefore influence $\dot{M}_{\text{acc}}$, but they have little impact on abundance ratios.

\subsection{Implementation}
\label{subsectionImplementation}

Bayes' theorem allows the parameters $\bm{\theta}$ of a model $\mathcal{M}$ to be inferred from data $\bm{x}$, taking account of prior knowledge or expectations $\pi(\bm{\theta})$, yielding posterior probability distributions $P(\bm{\theta})$. Uncertainties $\bm{\sigma}$ on the data are described by a likelihood function $\mathcal{L}(\bm{x})$, and a normalisation constant $\mathcal{Z}$ ensures that $P(\bm{\theta})$ integrates to unity. This is expressed in shorthand form as:

\begin{equation}
\label{equationBayesRuleShort}
    P(\bm{\theta}) = \frac{\mathcal{L}(\bm{x})\pi(\bm{\theta})}{\mathcal{Z}}
\end{equation}

\noindent
or more explicitly, in terms of probabilities:

\begin{equation}
\label{equationBayesRuleLong}
    p(\bm{\theta} | \bm{x},\mathcal{M}) = \frac{p(\bm{x}|\bm{\theta},\mathcal{M})p(\bm{\theta}|\mathcal{M})}{p(\bm{x}|\mathcal{M})}
\end{equation}

All models considered here evaluate Equations~\ref{equationAccretionT1} and~\ref{equationAccretionT2} for a given set of accreted abundances $\bm{X}_{\text{a}}$, to predict a set of model abundances $\bm{X}_{\text{m}}$. The models, their parameters, and their priors and are detailed below.

The data $\bm{x}$ are the observed photospheric abundances $\bm{X}$ determined in Section~\ref{sectionObservationsAndModelling}, for the $N$ metals detected at a given star. Uncertainties are assumed to be Gaussian, and are conservatively treated as independent as a covariance matrix cannot reliably be determined. The likelihood function $\mathcal{L}_{\text{m}}$ for an abundance measurement of element $k$ is:

\begin{equation}
\label{equationLikelihoodMeasurement}
    \mathcal{L}_{\text{m},k}=\frac{1}{\sigma_k\sqrt{2\pi}}\exp{\left[-\frac{1}{2}\left(\frac{X_k-X_{\text{m},k}}{\sigma_k}\right)^2\right]}
\end{equation}

In some cases, only upper limits are available, but these are still informative, so are included in the analysis. Upper limits are left-censored data, for which the survival function $S$ for a Gaussian provides a suitable likelihood \citep{Klein2003survival}:

\begin{equation}
\label{equationSurvivalFunction}
    S(X_k)=\frac{1}{2}\left[1-\text{erf}\left(\frac{X_k-X_{\text{m},k}}{\sqrt{2}\sigma_k}\right)\right]
\end{equation}

\begin{equation}
\label{equationLikelihoodUpperLimit}
    \mathcal{L}_{\text{ul},k}=1-S(X_k)
\end{equation}

%Cheat sheet: http://www.math.wm.edu/~leemis/chart/UDR/PDFs/Normal.pdf

To a good approximation, the probability distribution function equals unity for all values below the upper limit, transitioning to a Gaussian form above it. The uncertainty $\sigma$ is set at 0.3\,dex.

Equations~\ref{equationLikelihoodMeasurement} and~\ref{equationLikelihoodUpperLimit} are combined to give the full likelihood function, by setting a switching parameter $\delta=0$ for an upper limit, and $\delta=1$ for a measurement:

\begin{equation}
\label{equationLikelihoodCombined}
    \mathcal{L}=\prod_{k=1}^{N}\mathcal{L}_{\text{m},k}^{\delta_k}\mathcal{L}_{\text{ul},k}^{1-\delta_k}
\end{equation}

The Bayesian evidence $\mathcal{Z}$ in Equation~\ref{equationBayesRuleShort} can be neglected if only the distribution of $P(\bm{\theta})$ is of interest, as is often the case in parameter inference problems. However, it is essential for model comparison and averaging.  Equation~\ref{equationBayesRuleShort} is therefore evaluated for the models described in Section~\ref{sectionAnalysis} using \textsc{dynesty}, which estimates both $P(\bm{\theta})$ and $\mathcal{Z}$ via nested sampling \citep{Skilling2004,Skilling2006,Speagle2020dynesty}. The dynamic sampler is initialised with 1000 live points per dimension, and allowed to halt when the estimate of the evidence has stabilised to $\delta\ln{\mathcal{Z}}<0.01$. The computational cost of nested sampling increases with the prior volume explored, so boundaries are set on the priors so that implausible parameter values have a probability of zero. Those priors are described in Section~\ref{sectionAnalysis}, and quantified in Table~\ref{tablePriors}.

All models $\mathcal{M}$ share parameters $\bm{\theta}=(M_{\text{acc}}, t_1, t_2, \teff)$, and differ only in their compositions $\bm{X}_{\text{a}}$. However, the solar system comparison objects do not all have measurements of the same elements. While that does not preclude an analysis under an individual model, any model comparison or averaging requires that the data $\bm{X}$ for a given target be identical across all models $\mathcal{M}$. Therefore, a set of standard metals (O, Na, Mg, Al, Si, Ca, Ti, Cr, Mn, Fe, and Ni) is defined. Only comparison objects including all of those elements are considered, and any other elements represented in the target data are not considered. This does not prove overly restrictive: only a handful of comparison objects are unsuitable (e.g.~pallasite meteorites, core Earth, and comet 67P/C--G all lack a titanium abundance). Only upper limits for scandium and vanadium in the target data are ignored. A measurement of hydrogen is also required in a comparison composition, but where none is available it is assumed to be zero.

While in principle $\log{g}$ is an independent parameter, in practice it is highly correlated with $\teff$ \citep{Hoskin2020}. For computational efficiency, it is therefore varied in exact correlation with $\teff$, rather than being treated as a free parameter.

Sinking timescales and convection zone depths are determined using the Koester model atmosphere and envelope codes \citep{Koester2020atmospheres}. The atmosphere models, which serve as boundary conditions for the envelope integration, contain 15 metals fixed to bulk Earth abundances. They are varied using the calcium abundance as a metallicity parameter. Pre-calculated grids extending over a wide range of $\teff$, $\log{g}$, and $X_{\text{m,Ca}}$ are used, where convective overshoot is assumed to extend the mixing zone by one pressure scale height. Reference values for each target star are determined by interpolation, using the best-fit parameters obtained in Section~\ref{sectionObservationsAndModelling}, and are given in Table~\ref{tableTargetData}. During sampling, $\tau$ and $M_{\text{cvz}}$ are interpolated from the grid based on $\teff$ and $\log{g}$, with $X_{\text{Ca}}$ fixed at the observed value. Ratios between sinking times (and therefore inferred abundances) vary little within the uncertainties on $\teff$, but changes in their absolute values contribute to the uncertainty on accretion rate $\dot{M}_{\text{acc}}$ or total accreted mass $M_{\text{acc}}$.

\subsection{Model comparison and averaging}
\label{subsectionModelComparisonAveraging}

Models can be compared quantitatively using the Bayesian evidence $\mathcal{Z}$ from Equation~\ref{equationBayesRuleShort}. Their posterior distributions can also be averaged, marginalising over the models to yield a best-estimate of their common parameters. The procedures are summarised here, but see the review by \cite{Trotta2008} for more detail and examples.

It can be shown from Equation~\ref{equationBayesRuleLong} that for two models $\mathcal{M}_{\text{a}}$ and $\mathcal{M}_{\text{b}}$:

\begin{equation}
\label{equationPosteriorOdds}
    O_{\text{ab}}=\frac{p(\mathcal{M}_{\text{a}}|\bm{x})}{p(\mathcal{M}_{\text{b}}|\bm{x})}=B_{\text{ab}}\frac{p(\mathcal{M}_{\text{a}})}{p(\mathcal{M}_{\text{b}})}
\end{equation}

\noindent
where

\begin{equation}
\label{equationBayesFactor}
    B_{\text{ab}}=\frac{p(\bm{x}|\mathcal{M}_{\text{a}})}{p(\bm{x}|\mathcal{M}_{\text{b}})}=\frac{\mathcal{Z}_{\text{a}}}{\mathcal{Z}_{\text{b}}}
\end{equation}

In Equation~\ref{equationPosteriorOdds}, $O_{\text{ab}}$ is the ratio of the probabilities of each model, given the data, known as the posterior odds. This is the key quantity in model comparison and averaging. The Bayes factor $B_{\text{ab}}$ is often substituted for the posterior odds when equal prior probabilities $p(\mathcal{M})$ are assigned to each model, but that is not the case here. A neutral stance is adopted, where the solar system object and core--mantle--crust model families are assigned equal probabilities of $p=0.5$. As 31 individual compositions and two planets are considered in the respective model families, individual models are assigned $p=0.5/31$ or $p=0.5/2$ as appropriate. Confidence in a core--mantle--crust model therefore does not depend on the number of solar system compositions that are considered, and vice versa.

Two models are compared by inspecting their posterior odds. A value of $\ln{O_{\text{ab}}}\gtrsim5$ can be considered to be strong evidence that $\mathcal{M}_{\text{a}}$ is superior to $\mathcal{M}_{\text{b}}$ \citep{Trotta2008}. Interpretation of relative odds is subjective, so this threshold is as arbitrary as the commonly used (and approximately equivalent) 3-$\upsigma$ requirement for statistical significance, but it is adopted here to guide interpretation of the results. The quality of a model fit is also subjective, and the superiority of one over another says nothing about the suitability of either: Bayesian analysis always requires a model, so the one with the highest posterior odds may be merely the least bad choice from a set of poor options. Vetting of the best-fit solution is therefore advised.

As $\mathcal{Z}$ serves as a normalisation factor in Equation~\ref{equationBayesRuleShort}, its value depends on the priors $\pi(\bm{\theta})$. All models use the same priors for $M_{\text{acc}}$, $t_1$, and $t_2$, so comparisons between them are fair. However, the core--mantle--crust models introduces additional parameters, expanding the prior volume, so that $\mathcal{Z}$ must decrease to compensate. Thus, Bayesian model comparison applies Occam's razor: a more complex model must produce a significantly better fit than a simpler one if it is to be judged superior.

A second use for the Bayesian evidence is model averaging. A weighted average of the parameter inferences made under different models produces a best-estimate posterior distribution for their common parameters:

\begin{equation}
\label{equationModelAveraging}
    P_{\text{avg}}(\bm{\theta})=\sum_{i}p(\mathcal{M}_i|\bm{x})p(\bm{\theta}|\bm{x},\mathcal{M}_i)
\end{equation}

\noindent
where

\begin{equation}
\label{equationModeAveragingWeight}
    p(\mathcal{M}_i|\bm{x})=\frac{\mathcal{Z}_i p(\mathcal{M}_i)}{\sum_{i}\mathcal{Z}_i p(\mathcal{M}_i)}
\end{equation}

In practice, the averaged posterior distributions are constructed by resampling and combining the outputs of the original nested sampling runs for each model, appropriately weighted. Where one model is clearly superior, its posterior will be very similar to the averaged posterior, but where there are multiple competing models, averaging effectively marginalises over them, and avoids the need to pick a winner.

Given the averaged posterior values of $M_{\text{acc}}$, $t_1$, and $t_2$, the size and composition of the accreted object can be estimated by working back from the photospheric abundances using Equations~\ref{equationAccretionT1} and~\ref{equationAccretionT2}. Again, it is reiterated that if none of the models can account well for the data, their averaged posteriors will represent merely the least-bad estimate of the parameters.

%%%%%%%%%%%%%%%%%%%%%%%%%%%%%%%%%%%%%%%%%%%%%%%%%%

% Don't change these lines
\bsp	% typesetting comment
\label{lastpage}
\end{document}